\documentclass[journal]{IEEEtran}
\usepackage{amsmath,amsfonts}
\usepackage{array}
\usepackage{textcomp}
\usepackage{stfloats}
\usepackage{url}
\usepackage{verbatim}
\usepackage{graphicx}
\usepackage{colortbl}
\usepackage{arydshln}
\usepackage{cite}
\usepackage{orcidlink}
\usepackage{subfigure}
\usepackage{booktabs}
\usepackage{multirow}
\usepackage{bm}
\usepackage{epstopdf}
\usepackage{xcolor}
\usepackage{mathrsfs}
\usepackage[ruled]{algorithm2e}
\usepackage{amssymb}
\usepackage[figuresright]{rotating}
\usepackage{graphicx} 
\usepackage{subcaption}
\usepackage{color}
\usepackage{hyperref} 
\usepackage{authblk}

\begin{document}
\newenvironment{proof}{{\indent \it Proof:}}{\hfill $\blacksquare$\par}
\title{QuantFactor REINFORCE: Mining Steady Formulaic Alpha Factors with Variance-bounded REINFORCE}

\author{Junjie Zhao$^{\orcidlink{0000-0003-0718-4183}}$,
Chengxi Zhang$^{\orcidlink{0009-0009-5876-5632}}$,
Min Qin,
Peng Yang$^{\orcidlink{0000-0001-5333-6155}}$,~\IEEEmembership{Senior Member,~IEEE}
        
\thanks{This paper was produced by the IEEE Publication Technology Group. They are in Piscataway, NJ.}
\thanks{Manuscript received xxx, xxx. This work was supported by National Natural Science Foundation of China (Grants 62272210, 62250710682, and 62331014).\textit{(Corresponding
author: Peng Yang.)}

Junjie Zhao and Peng Yang are with the Guangdong Provincial Key Laboratory of Brain-Inspired Intelligent Computation, Department of Computer Science and Engineering, Department of Statistics and Data Science, Southern University of Science and Technology, Shenzhen 518055, China (e-mail: zhaojj2024@mail.sustech.edu.cn; yangp@sustech.edu.cn). Junjie Zhao is also with the Shenzhen Yujin Hedge Fund Company Limited, Shenzhen 518033, China (e-mail: junjie.zhao@yujinamc.com).

Chengxi Zhang is with the Department of Modern Physics, University of Science and Technology of China, Hefei 230026, China (e-mail: zhangchengxi@mail.ustc.edu.cn). 

Min Qin is with the Shenzhen Yujin Hedge Fund Company Limited, Shenzhen 518033, China (e-mail: min@yujinamc.com).}}



\maketitle

\begin{abstract}
Alpha factor mining aims to discover investment signals from the historical financial market data, which can be used to predict asset returns and gain excess profits. 
Powerful deep learning methods for alpha factor mining lack interpretability, making them unacceptable in the risk-sensitive real markets. 
Formulaic alpha factors are preferred for their interpretability, while the search space is complex and powerful explorative methods are urged. Recently, a promising framework is proposed for generating formulaic alpha factors using deep reinforcement learning, and quickly gained research focuses from both academia and industries. This paper first argues that the originally employed policy training method, i.e., Proximal Policy Optimization (PPO), faces several important issues in the context of alpha factors mining.
Herein, a novel reinforcement learning algorithm based on the well-known REINFORCE algorithm is proposed. REINFORCE employs Monte Carlo sampling to estimate the policy gradient—yielding unbiased but high variance estimates. The minimal environmental variability inherent in the underlying state transition function, which adheres to the Dirac distribution, can help alleviate this high variance issue, making REINFORCE algorithm more appropriate than PPO. A new dedicated baseline is designed to theoretically reduce the commonly suffered high variance of REINFORCE. Moreover, the information ratio is introduced as a reward shaping mechanism to encourage the generation of steady alpha factors that can better adapt to changes in market volatility. 
Evaluations on real assets data indicate the proposed algorithm boosts correlation with returns by 3.83\%, and a stronger ability to obtain excess returns compared to the latest alpha factors mining methods, which meets the theoretical results well.

\end{abstract}


\section{Introduction}

\IEEEPARstart{C}{onstructing} a superior portfolio involves distinguishing noise from the huge raw data in financial markets and discovering signals to balance risk and return, which is widely regarded as a signal processing problem\cite{shi2024saoftrl,zhao2019optimal,zhao2018mean}. In the realm of computational finance, it has become nearly a conventional practice to convert raw historical asset information into alpha factor values\cite{zhang2019deeplob}, which serve as indicative signals of market trends for portfolio management\cite{qian2007quantitative}. The functions that generate these signals are known as alpha factors, which can be expressed in two distinct forms: the deep model and the formula\cite{qian2007quantitative}. Mining high-quality alpha factors has become a trendy topic among investors and researchers, due to their relationship with excess investment returns\cite{10494678}.

Although alpha factors using end-to-end deep models are generally more expressive, they are in nature black-box and poorly interpretable. Thus, when the performance of these black-box models deteriorates unexpectedly, it is often less likely for human experts to adjust the models accordingly\cite{hassija2024interpreting}. Therefore, due to the need for risk control, it is difficult to involve these black-box factors in practical trading.

Comparatively, alpha factors represented in formulaic forms enjoy much better interpretability and thus are favored by market participants. 
The most intuitive method for automatically mining formulaic factors is the tree-based models, which mutate expression trees to generate new alpha factors\cite{zhang2023openfe, zhu2022application, li2022research}. Another commonly used method is the genetic programming (GP), which generates the formulaic alpha factor expressions through iterative simulations of genetic variation and natural selection in biological evolution\cite{yang2024reducing}. 
Unfortunately, both methods still face certain challenges. Tree-based models, while easy to understand and implement, may encounter performance bottlenecks when dealing with non-linear relationships and high-dimensional data\cite{grinsztajn2022tree}. Genetic programming, on the other hand, can deal with broader types of expressions by properly defining the search space, including complex nonlinear and non-parametric factor expressions, but it usually fails to explore the search space of large-scale expressions. Additionally, it is usually computationally expensive\cite{ismail2023prediction}.

Recently, Yu et al. \cite{yu2023generating} proposes a promising framework named AlphaGen for mining formulaic alpha factors with Reinforcement Learning (RL), trying to bridge both deep learning and formulaic based methods. It employs Markov Decision Processes (MDPs)\cite{puterman2014markov} to simulate the generation process of formulaic alpha factors and trains policies to directly generate a set of collaborative formulaic alpha factors using RL. It essentially aims at finding formulaic alpha factors. With the help of RL, it also overcomes the limitations of the explorative search ability suffered by traditional tree models and genetic programming\cite{yu2023generating}.

The key to this framework lies in using the Reverse Polish Notation (RPN) to convert formulaic alpha factors into a sequence of tokens. Tokens include various calculation operators, the original asset features, financial constants, and other elements that can constitute a formula. The action in the MDPs is represented as a new token to be added to the sequence, and the state of the MDPs is the current sequence composed of the previous tokens (actions). The reward can be set as any well-established performance indicators of alpha factors. The remaining problem is how to learn the policy effectively. There are generally two ways to estimate the policy gradient: Temporal Difference Sampling (e.g., the actor-critic architecture) and Monte Carlo sampling (e.g., without using the critic network). The Temporal Difference method may be biased because it adopts the estimated values of the critic network rather than actual rewards, but its variance can be made small as it updates the networks after each action, thereby reducing the impact of randomness. The Monte Carlo method is unbiased as it receives the actual rewards. However it suffers from high variance since it relies on complete episodes, i.e., the fully constructed formulas. 

In \cite{yu2023generating}, the actor-critic architecture is adopted, with an actor network as the policy and a critic network for reducing the variance in the training process. Naturally, the seminal Proximal Policy Optimization (PPO) \cite{schulman2017proximal} was chosen to train the networks. This paper proposes that the actor-critic architecture and PPO may not be suitable for this factor mining framework. Specifically, the aforementioned alpha factors mining framework has trajectory feedback, i.e., non-zero reward can only be obtained after completing a full trajectory \cite{yang2021parallel,yang2022evolutionary}. This makes the critic network of PPO difficult to extract useful training signals from the intermediate states, leading to inevitable biases when attempting to estimate the value of the state and thus slow converging speed. Furthermore, the critic network typically has the same scale with the policy network, meaning that two large networks need to be updated in each iteration of PPO. This not only makes the training harder, but leads to doubled parameter updating time\cite{li2023remax}.


Based on the above discussions, this paper proposes a new RL-based method for formulaic alpha factors mining, named QuantFactor REINFORCE (QFR). QFR abandons the critic network of AlphaGen and train the policy using the REINFORCE\cite{williams1987reinforcement}, a seminal Monte Carlo method for policy gradient estimation. In this regard, the biases of PPO can be vanished. The high variance of REINFORCE is mitigated in two aspects. For the variance caused by the environment, we observe that the underlying MDPs has deterministic transitions, which satisfies the Dirac distribution. That is, once a new action is selected, the new state is uniquely determined by the current sequence and the newly generated token within the RPN representation. This ensures that REINFORCE will not be degenerated by the environmental variance. For the variance caused by the policy gradient estimation, QFR employs a greedy policy to generate a novel baseline, which has been theoretically shown to bound the variance and leads to a lower variance than the original REINFORCE.
Additionally, we introduce the Information Ratio (IR) for reward shaping to better balance returns and risks. As a result, QFR enjoys faster convergence speed to the optimal policy and produces steadier alpha factor signals across various market volatility.



To evaluate our proposed QFR algorithm, we conducted extensive experiments on multiple real-world asset datasets. Our experimental results show that the set of formulaic alpha factors generated by QFR outperforms those generated by previous methods. It is also verified that theoretical results properly predict trends in the experimental results, even encountering distinct market volatility.

The \textbf{main contributions} of the paper are as follows:

\begin{itemize}
    %
    \item We propose a more stable and efficient RL algorithm for mining formulaic alpha factors. Unlike prior research that employs the actor-critic framework, we argue to discard the critic network due to the trajectory feedback and theoretically show that the high variance due to the absence of the critic network can be significantly mitigated.
    
    \item Two new components are proposed for RL-based formulaic alpha factors mining. A novel baseline that employs a greedy policy is proposed to reduce the variance of the policy gradient estimation. Additionally, IR is introduced as a reward shaping mechanism to balance returns and risks. We also provide a set of theoretical results, including an analysis of the training variance under state transition functions with various distributions, the derivation of the upper bound of the variance after introducing the baseline, and the proof that the variance decreases compared to REINFORCE.
    
    \item Extensive experiments on multiple real-world asset datasets show that the proposed algorithm outperforms the RL algorithm used in work \cite{yu2023generating}, as well as various tree-based and heuristic algorithms. Compared to the previous state-of-the-art algorithm, it improves the correlation with asset returns by 3.83\%, while also demonstrating a stronger ability to generate excess profits. Additionally, the experimental results closely align with the theoretical results.

\end{itemize}

The rest of this paper is organized as follows. The formulaic alpha factors for predicting asset prices are introduced and the corresponding MDPs is formulated in Section II. Section III details the proposed QFR algorithm and provides thorough theoretical analyses. Numerical results are presented in Section IV to show the performance of the QFR. Finally, Section V concludes this paper.

We use the following notation: vectors are bold lower case $\mathbf{x}$; matrices are bold upper case $\mathbf{A}$; sets are in calligraphic font $\mathcal{S}$; and scalars are non-bold $\alpha$.

\section{Related Works}

This section provides a brief review of the related work on the automatic mining methods for alpha factors and the theory of REINFORCE algorithm.\\[2.5pt]

\subsection{Automatic Mining for Alpha Factors}
Alpha factors are generally represented in the form of deep models or formulas. Alpha factors using end-to-end deep models are more complex and usually trained with supervised learning\cite{zhang2019deeplob}, utilizing Multilayer Perceptron (MLP)\cite{wang2023accurate,liu2025multiscale,huang2018machine,huang2018machine1,huang2020q} or sequential models like Long Short Term Memory (LSTM)\cite{graves2012long} and Transformer\cite{vaswani2017attention} to extract features embedded in the historical data of assets. Recently, reinforcement learning has attracted much attention in the context of computational finance and fintech. It becomes a key technology in alpha factor mining\cite{yu2023generating}, investment portfolio optimization\cite{huang2020deep}, and risk management design\cite{li2023combining}, thanks to its advantages in handling non-linearity, high-dimensional data, and dynamic environments\cite{zhao2024mimic}. By modeling market characteristics as states, maker or taker orders as actions, and profit and loss as rewards, RL can also be used to train deep policy models that represents alpha factors\cite{liu2020finrl, he2023multi, SHAVANDI2022118124}. 

On the other hand, alpha factors represented in formulaic forms have much better interpretability and thus are favored by market participants. In the past, these formulaic alpha factors were constructed by human experts using their domain knowledge and experience, often embodying clear economic principles. For example, Kakushadze \cite{kakushadze2016101} presents 101 formulaic alpha factors tested in the US asset market. However, the alpha factor mining process relying on human experts suffers from multiple drawbacks such as strong subjectivity\cite{ye2021prediction}, time-consuming\cite{ye2021prediction}, insufficient risk control\cite{zhang2021pyramid}, and high costs\cite{zhao2022objective}. To address these issues, algorithms for automatically mining formulaic alpha factors have been proposed \cite{yu2023generating}, such as tree models represented by Gradient Boosting Decision Tree (GBDT)\cite{zhang2023openfe}, eXtreme Gradient Boosting (XGBoost)\cite{zhu2022application}, and Light Gradient Boosting Machine (LightGBM)\cite{li2022research}, as well as heuristic algorithms represented by GP\cite{zhang2020autoalpha}. These algorithms can quickly discover numerous new formulaic alpha factors without requiring the domain knowledge or experience of human experts. They offer performance comparable to more complex deep learning-based alpha factors while maintaining relatively high interpretability.

The work in \cite{yu2023generating} is the first to employ RL for formulaic alpha factors mining. It employs MDPs to simulate the generation process of formulaic alpha factors and trains policies to directly generate a set of collaborative formulaic alpha factors using RL. Compared to works \cite{liu2020finrl, he2023multi, SHAVANDI2022118124} that directly use RL to build trading agents, this framework uses historical quantitative and price data of assets as input to find a set of formulaic alpha factors with strong interpretability, avoiding the black-box problem and overcoming the limitations of \cite{zhang2023openfe, zhu2022application, li2022research, zhang2020autoalpha} in independently mining individual formulaic factors, such as homogenization among factors, lack of synergy, and difficulty in adapting to dynamic market changes. The characteristics of RL, supervised learning, tree models, and heuristic algorithms when applied to mining alpha factors are shown in Table \ref{Character of Alpha Mining Algorithms}.

\begin{table*}[thb]
\centering
\caption{Comparison of Various Factor Mining Algorithms}
\begin{tabular}{ccccc}
\toprule
                                & Reinforcement Learning                                                                      & Supervised Learning                             & Tree Models                                                                                 & Genetic Programming                         \\ \hline
Representative Algorithms       & AlphaGen\cite{yu2023generating}, Finrl\cite{liu2020finrl} & DeepLOB\cite{zhang2019deeplob} & OpenFE\cite{zhang2023openfe}, Alpha360\cite{yang2020qlib} & GP\cite{zhang2020autoalpha} \\
The Number of Factors Generated & From $10^0$ to $10^2$                                                                       & From $10^0$ to $10^1$                           & More than $10^3$                                                                            & More than $10^4$                             \\
Interpretability                & Varied                                                                                      & Worst                                           & Good                                                                                        & Best                                         \\
In-sample Performance           & Good                                                                                        & Best                                            & Good                                                                                        & Good                                         \\
Operator Requirements           & None                                                                                        & Strict (differentiable)                         & None                                                                                        & None                                         \\
Convergence Efficiency          & Fast                                                                                        & Fastest                                         & Slow                                                                                        & Slowest                                      \\ \bottomrule
\end{tabular}
\label{Character of Alpha Mining Algorithms}
\end{table*}

\subsection{The REINFORCE algorithms}
Williams is the first to introduce the REINFORCE algorithm in his works\cite{williams1987reinforcement}. The algorithm is straightforward and versatile, suitable for a wide range of tasks that can be modeled as MDPs\cite{williams1992simple}. However, in MDPs with stochastic state transitions and immediate rewards, it often underperforms compared to actor-critic methods\cite{konda1999actor}. Actor-critic methods, which utilizes value-based techniques to decrease variance and integrate temporal-difference learning for dynamic programming principles, are generally favored. Consequently, the REINFORCE algorithm has not gained widespread favor in the RL community. However, our work demonstrates that the REINFORCE algorithm can be suitable to MDPs for mining formulaic factors as long as a proper baseline is used to reduce the variance. 

A series of studies have explored the incorporation of a baseline value in the REINFORCE algorithm. To our best knowledge, \cite{dayan1991reinforcement} is the pioneer in demonstrating that employing the expected reward as a baseline does not universally decrease variance, and he introduced an optimal baseline for a basic 2-armed bandit scenario to address the issue.  Additionally, the regret of REINFORCE algorithm was studied\cite{zhang2021sample}. Recently, The work in \cite{ mei2021leveraging } delved into the intricacies of multi-armed bandit optimization, revealing that the expected, as opposed to the stochastic, gradient ascent in REINFORCE can lead to a globally optimal solution. Building on this, the work was extended to the function of the baseline value in the natural policy gradient\cite{kakade2002approximately}, concluding that variance reduction is not critical for natural policy gradient \cite{mei2022role}. More recently, the work in \cite{li2023remax} find that the REINFORCE is suitable for reinforcement learning from human feedback in large language models, and proposed a baseline value serves as a kind of normalization by comparing the rewards of a random response with those of the greedy response.

\color{black}{\section{Problem Formulation and Preliminaries}}
This section first introduces the definition of formulaic alpha factors and their RPN sequences. Next, the MDPs for mining formulaic alpha factors modeled by Yu et al. \cite{yu2023generating} are detailed. Lastly, the two seminal methods for solving the modeled MDPs, i.e., PPO and REINFORCE, are compared to motivate the proposed QFR algorithm.\\[2.5pt]

\subsection{Alpha Factors for Predicting Asset Prices}
\label{Alpha Factors for Predicting Asset Prices}
Consider a real market with \( n \) assets over \( L \) trading days. On each trading day \( l \in \{1, 2, \cdots, L\} \), each asset \( i \) corresponds to a feature vector \( \mathbf{X}_{li}\in \mathbb{R}^{m\times d}\). This vector consists of \( m \) raw market features, such as open, high, low, close, and volume values, over the recent \( d \) days. Here, \(\mathbf{x}_{lij} \in \mathbb{R}^{d \times 1}\) denotes the sequence of values for the \( j \)-th raw feature over these \( d \) days.

Next, we define an alpha factor function \( f \), which transforms the feature matrix for all $n$ assets on the $l$-th trading day, represented as \(\mathbf{X}_l = [\mathbf{X}_{l1}, \mathbf{X}_{l2}, \cdots, \mathbf{X}_{ln}]^\mathsf{T} \in \mathbb{R}^{n \times m \times d}\), into alpha factor values \(\mathbf{z}_l = f(\mathbf{X}_l)\in \mathbb{R}^{n \times 1}\). Specifically, \(\mathbf{z}_l\) holds the alpha factor values for all assets on the \( l \)-th trading day. The real asset feature dataset over $L$ days is denoted as $\mathcal{X} = \{\mathbf{X}_l\}$.

After the policy model outputs a new formulaic alpha factor, this alpha will be added to an alpha factors pool $\mathcal{F}$, say $\mathcal{F}=\{f_1,f_2,...,f_K\}$ with $K$ factors. 
The linear combination model is adopted to compute the asset price predicted by the alpha factor pool\cite{yu2023generating}.
Specifically, suppose each $k$-th alpha factor ($1 \leq k\leq K$) is associated with a weight $w_k$, and the asset price is calculated as $\mathbf{z}_l^{\prime}=\sum^K_{k=1} w_kf_k(\mathbf{X}_l)$. 
In other words, the mining of alpha factors is to search a set of alpha factors and use them in combination.
The weight of each factor indicates its exposure on the assets and is optimized using gradient descent.
The loss function of learning the weights vector $\mathbf{\omega} \in \mathbb{R}^{K\times 1}$ is defined as the mean squared error (MSE) between the model output and the ground-truth asset prices $\mathcal{Y} = \{\mathbf{y}_l\}$ with \( l \in \{1, 2, \cdots, L\} \) and \( \mathbf{y}_l \in \mathbb{R}^{n\times 1} \):
\begin{equation}
    L(\mathbf{\omega})=\frac{1}{L} \sum_{l=1}^{L}\left\|\mathbf{z}_l^{\prime}-\mathbf{y}_l \right\|^{2}.    
    \label{combination model loss}
\end{equation}

\noindent Due to significant differences in the scales of different alpha factor values, they are first normalized to have a mean of $0$ and a maximum value of $1$. If the number of factors in the pool exceeds a certain threshold, the alpha factor with the smallest weight and its corresponding weight will be discarded.


\subsection{Formulaic Alpha Factors}
\label{Formulaic Alpha Factors}

\begin{figure}[tbh]
\centering
\includegraphics[width=0.49\textwidth]{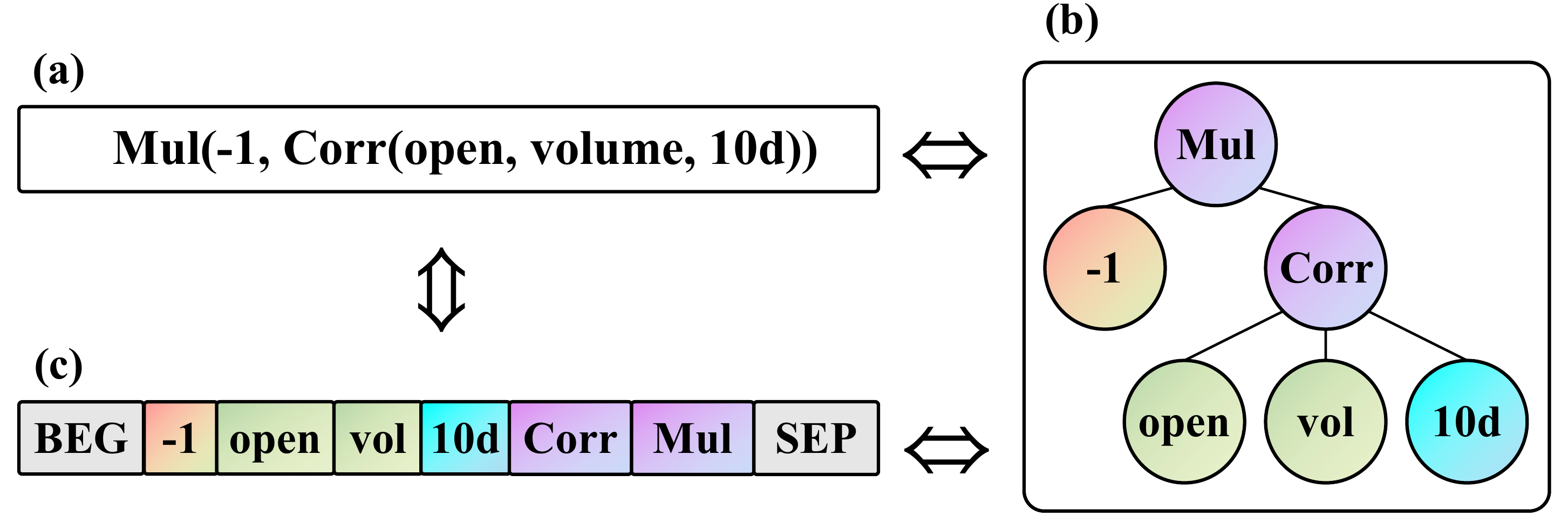}
\caption{Three uniquely interchangeable forms of an alpha factor: (a) formulaic expression; (b) tree structure; (c) RPN sequence.}
\label{Figure1}
\end{figure}

Formulaic alpha factors are mathematical expressions that can be represented using RPN, which is a sequence of tokens. Tokens include various operators, the original volume-price features, fundamental features, time deltas, constants, and sequence indicators. The operators include elementary functions that operate on single-day data, known as cross-sectional operators (e.g., Abs($x$) for the absolute value $|x|$, Log($x$) for the natural logarithm log($x$)), as well as functions that operate on a series of daily data, known as time-series operators (e.g., Ref($x,l$) for the expression $x$ evaluated at $l$ days before the current day, where $l$ denotes a time token, such as 10d (10 days)). The Begin (BEG) token and Separator (SEP) token of the RPN representation are used to mark the beginning and end of the sequence. Table \ref{Examples of the Tokens} illustrates a selection of these tokens as examples.
Such formulas can naturally be represented by an expression tree, with each non-leaf node representing an operator, and the children of a node representing the original volume-price features, fundamental features, time deltas, and constants being operated on. Each such expression has a unique post-order traversal, using RPN. An example of a formulaic alpha expression, together with its corresponding tree and RPN sequence, is shown in Fig. \ref{Figure1}. 

\begin{table}[bp]
\centering
\caption{Examples of Formulaic Tokens}
\begin{tabular*}{0.35\textwidth}{@{\extracolsep{\fill}}cc}
\toprule
Token Types         & Token Instances              \\ \hline
Operators           & Abs($x$), Log($x$), Ref($x,l$) \\
Features            & open, high, low, close       \\
Time Deltas         & 10d, 20d, 50d                 \\
Constants           & -10, -5, -0.01, 0.01, 5, 10  \\
Sequence Indicators & BEG, SEP                     \\ \bottomrule
\end{tabular*}
\label{Examples of the Tokens}
\end{table}

To measure the effectiveness of a formulaic alpha factor, the Pearson correlation coefficient between the ground-truth asset price \( \mathbf{y}_l \) and the combination factor value \( \mathbf{z}_l^{\prime}\), also known as the Information Coefficient (IC), is commonly employed as the performance indicators which calculates as follows:

\begin{equation} IC\left(\mathbf{z}^{\prime}_l, \mathbf{y}_l\right) = \frac{\text{Cov}(\mathbf{z}^{\prime}_l, \mathbf{y}_l)}{\sigma_{\mathbf{z}^{\prime}_l} \sigma_{\mathbf{y}_l}}. 
\label{IC defination}
\end{equation}

\noindent where $\mathrm{Cov}(\cdot,\cdot)$ indicates the covariance matrix between two vectors, and $\sigma_{\cdot}$ means the standard deviation of a vector. The IC serves as a critical metric in quantitative finance, where higher values indicate stronger predictive power. This directly translates to portfolio enhancement through more informed capital allocation: factors with elevated IC enable investors to overweight assets with higher expected returns while underweighting underperformers.
The averaged IC values over all $L$ trading days are denoted as \(\overline{IC}=\mathbb{E}_{l}\left[IC\left(\mathbf{z}^{\prime}_l, \mathbf{y}_l\right)\right]=\frac{1}{L} \sum^L_{l=1} IC\left(\mathbf{z}^{\prime}_l, \mathbf{y}_l\right)\).

\subsection{MDPs for Mining Formulaic Alpha Factors}
\label{Alpha Factor Mining Processes}

The process of generating a sequence of tokens that can equivalently represent a formulaic alpha factor with RPN is modeled as an MDP\cite{yu2023generating}, which can be described in the classic tuple of $\{\mathcal{S}, \mathcal{A}, P, r\}$. Specifically, $\mathcal{A}$ denotes the finite action space, consisting of a finite set of candidate tokens as actions $a$. $\mathcal{S}$ represents the finite state space, where each state at the $t$-th time step corresponds to the sequence of selected tokens, representing the currently generated part of the formulaic expression in RPN, denoted as \(\mathbf{s}_t = \mathbf{a}_{1:t-1} = [a_1, a_2, \cdots, a_{t-1}]^\mathsf{T}\). Our goal is to train a parameterized policy $\pi_\mathbf{\theta}:\mathcal{S} \rightarrow \mathcal{A}$, which generates optimal formulaic alpha factors by iteratively selecting from the candidate tokens. The selecting process can thus be modeled as $a_{t} \sim \pi_\mathbf{\theta}\left(\cdot \mid \mathbf{a}_{1: t-1}\right)$, where the action $a_t$ is the next token following the currently generated part of the expression $\mathbf{a}_{1:t-1}$ in RPN sequence. 

The transition function $P$ defines the state transitions. 
When \(\mathbf{a}_{1:t-1}\) and \(a_t\) are already known, then \(\mathbf{s}_{t+1} = \mathbf{a}_{1:t}\) is uniquely determined, which means that the state transition function $P$ satisfies the Dirac distribution:
\begin{equation}
    P\left(\mathbf{s}_{t+1} \mid \mathbf{a}_{1:t}\right)=\left\{\begin{array}{ll}1 & \text { if } \mathbf{s}_{t+1}=\mathbf{a}_{1:t} \\ 0 & \text { otherwise. }\end{array}\right.
\end{equation}
An legal formulaic always starts with the begin (token) BEG, followed by any token selected from $\mathcal{A}$, and ends when the separator token (SEP) is selected or the maximum length is reached. Obviously, any generated sequence cannot be guaranteed to be a legal RPN sequence, therefore \cite{yu2023generating} only allow specific actions to be selected in certain states to ensure the correct format of the RPN sequence. For more details about these settings, please refer to \cite{yu2023generating}.

The reward function \(r: \mathcal{S} \times \mathcal{A} \rightarrow \mathbb{R}\) assigns values to the state-action pairs and is set to \(r\left(\mathbf{a}_{1: T}\right)=\overline{IC}\) in \cite{yu2023generating}.
The optimization objective in this MDP is to learn a policy $\pi_\mathbf{\theta}$ that maximizes the expected cumulative reward over time:

\begin{equation}
    J(\theta) = \mathbb{E}_{\mathbf{a}_{1: T} \sim \pi_{\mathbf{\theta}}}\left[ r\left(\mathbf{a}_{1: T}\right) \right].
    \label{original objective function}
\end{equation}

\noindent It is clear that non-zero rewards are only received at the final $T$-th step, which evaluates the quality of a complete formulaic factor expression, not individual tokens:
\begin{equation}
    r\left(\mathbf{s}_{t}, a_{t}\right)=\left\{\begin{array}{ll}0 & \text { if } t \neq T \\ r\left(\mathbf{a}_{1: T}\right) & \text { otherwise. }\end{array}\right.
\end{equation}

\noindent Notably, this MDP only models the formulaic alpha factors generation process, and the environment here specifically refers to the QFR's RL environment, rather than the financial market's random behavior. Its deterministic feature is manually designed and holds true regardless of the specific characteristics of the financial market.


\subsection{Comparing REINFORCE with PPO}
Gradient ascent is a typical way of learning the policy $\pi_\mathbf{\theta}$ by iteratively optimizing \(\theta\): \( \mathbf{\theta}_{k+1} \leftarrow \mathbf{\theta}_{k}+\eta_{k} \cdot g\left(\mathbf{\theta}_{k}\right)\), where $g(\mathbf{\theta}_{k})$ is the policy gradient at the $k$-th iteration and \(\eta_{k}\) represents the corresponding learning rate. The general policy gradient $g(\mathbf{\theta})$ is calculated as follows:
\begin{align}
	g(\mathbf{\theta}) & = \nabla_{\mathbf{\theta}} \mathbb{E}_{\tau \sim p_{\mathbf{\theta}}(\tau)}[R(\tau)] \nonumber \\ 
    & = \sum_{\tau} \nabla_{\theta} p_{\mathbf{\theta}}(\tau) R(\tau)\nonumber\\
    & =\mathbb{E}_{\tau \sim p_{\mathbf{\theta}}(\tau)}\left[\nabla_{\mathbf{\theta}} \log p_{\mathbf{\theta}}(\tau) R(\tau)\right] \nonumber \\ 
    & =\mathbb{E}_{\tau \sim p_{\mathbf{\theta}}(\tau)}\left[\sum_{t=0}^{T} \nabla_{\mathbf{\theta}} \log \pi_{\mathbf{\theta}}\left(a_{t} \mid \mathbf{a}_{1: t-1}\right) R(\tau)\right] \nonumber\\
    & =\mathbb{E}_{\mathbf{a}_{1: t} \sim \pi_{\mathbf{\theta}}}\left[\sum_{t=0}^{T} s_{\mathbf{\theta}}\left(\mathbf{a}_{1: t}\right) r\left(\mathbf{a}_{1: T}\right)\right],
    \label{Policy Gradient}
\end{align}
where \( \tau \) represents a trajectory, \( p_{\mathbf{\theta}}(\tau) \) represents the probability of trajectory \( \tau \) sampled by policy \( \pi_{\mathbf{\theta}} \). In the above MDP, the current policy is \(\pi_{\mathbf{\theta}}\left(a_{t} \mid \mathbf{a}_{1: t-1}\right)\), and thus we have \( p_{\mathbf{\theta}}(\tau)=p\left(a_{0}\right) \prod_{t=0}^{T} \pi_{\mathbf{\theta}}\left(a_{t} \mid \mathbf{a}_{1: t-1}\right) \).
 \( R(\tau) \) represents the cumulative reward of trajectory \( \tau \). Since IC can only be calculated after the complete expression is generated, only the reward for the final step is non-zero. Therefore, \(R(\tau) = r\left(\mathbf{a}_{1: T}\right)\). Let the score function be denoted as \(s_{\mathbf{\theta}}\left(\mathbf{a}_{1: t}\right)=\nabla_{\mathbf{\theta}} \log \pi_{\mathbf{\theta}}\left(a_{t} \mid \mathbf{a}_{1: t-1}\right)
\). 


Here \(r\left(\mathbf{a}_{1: t}\right)\) is used as the metric of the effectiveness of the current policy. It can also be replaced by the following five metrics: \(\sum_{t=0}^{\infty} r_t\) (total reward of the trajectory)\cite{sutton2018reinforcement}, \(\sum_{t^{\prime}=t}^{\infty} r_{t^{\prime}}\) (reward following action \(a_t\))\cite{sutton2018reinforcement}, \(\sum_{t^{\prime}=t}^{\infty} r_{t^{\prime}} - b(\mathbf{s}_t)\) (baselined version of the previous formula)\cite{sutton2018reinforcement}, \(Q^\pi(\mathbf{s}_t, a_t)\) (state-action value function)\cite{watkins1992q}, \(r_t + V^\pi(\mathbf{s}_{t+1}) - V^\pi(\mathbf{s}_t)\) (TD residual)\cite{schulman2015high}, and \(A\left(\mathbf{s}_{t}, a_{t}\right) = Q^\pi(\mathbf{s}_t, a_t) - V^\pi(\mathbf{s}_t)\) (advantage function)\cite{schulman2015high}. PPO \cite{schulman2017proximal} adopts the advantage function to measure the effectiveness of the policy under the actor-critic framework of TD\cite{schulman2015high}, and was employed in \cite{yu2023generating}. Both the actor network (the policy) and a critic network (the value network) are trained at the same time.
Generally, the merits of TD methods over the Monte Carlo algorithms\cite{swiechowski2023monte} to update network weights is that data from each step can be used for training, increasing the sample efficiency.

Unfortunately, the MDP considered in this work is with only trajectory feedback, i.e., non-zero rewards are only received at the final step. This means that the advantage function, as well as PPO, becomes less effective, since the critic network is hard to train and involves high biases. Comparatively, Monte Carlo algorithms without critic networks such as REINFORCE are better suited to. Generally, the REINFORCE calculates the policy gradient based on (\ref{Policy Gradient}) using the Monte Carlo method:
\begin{equation}
    \widehat{g}(\mathbf{\theta})=\frac{1}{N} \sum_{i=1}^{N} \sum_{t=1}^{T} s_{\mathbf{\theta}}\left(\mathbf{a}_{1: t}^{i}\right) r\left(\mathbf{a}_{1: T}^{i}\right),
\end{equation}
where \( a_{t}^{i} \sim \pi_{\mathbf{\theta}}\left(\cdot \mid \mathbf{a}_{1: t-1}^{i}\right) \) for all \(i=1, \ldots, N\). It is essentially a likelihood maximization weighted by rewards using samples extracted from rollouts. And for \(r\left(\mathbf{a}_{1: t}\right)\), REINFORCE normally adopts one of the first three metrics aforementioned and discards the critic network, thus is an unbiasesd estimation of policy gradient. 
We will delve into the theoretic details in Section \ref{The Theoretical Analysis}.

\section{Steadier RL for Alpha Mining}
In light of the potentially high biases caused by PPO, this section introduces a novel REINFORCE-based method for formulaic alpha factor mining, termed QuantFactor REINFORCE (QFR). Due to the Monte Carlo nature, QFR estimates unbiased policy gradient. Considering the high variance of REINFORCE methods, an effective greedy baseline is proposed, theoretically grounded by an upper bound on the variance. We demonstrate that QFR exhibits reduced variance compared to REINFORCE. The proof that QFR, when applied to MDPs with deterministic transition function, exhibits the lowest variance compared to those with probabilistic transition function, is also provided. These theoretical analysis supports a steady mining process of formulaic alpha factors. Furthermore, the Information Ratio (IR) is introduced as a new reward shaping mechanism to further promote the steadiness. 

\begin{figure}[!t]
\centering
\includegraphics[width=0.43\textwidth]{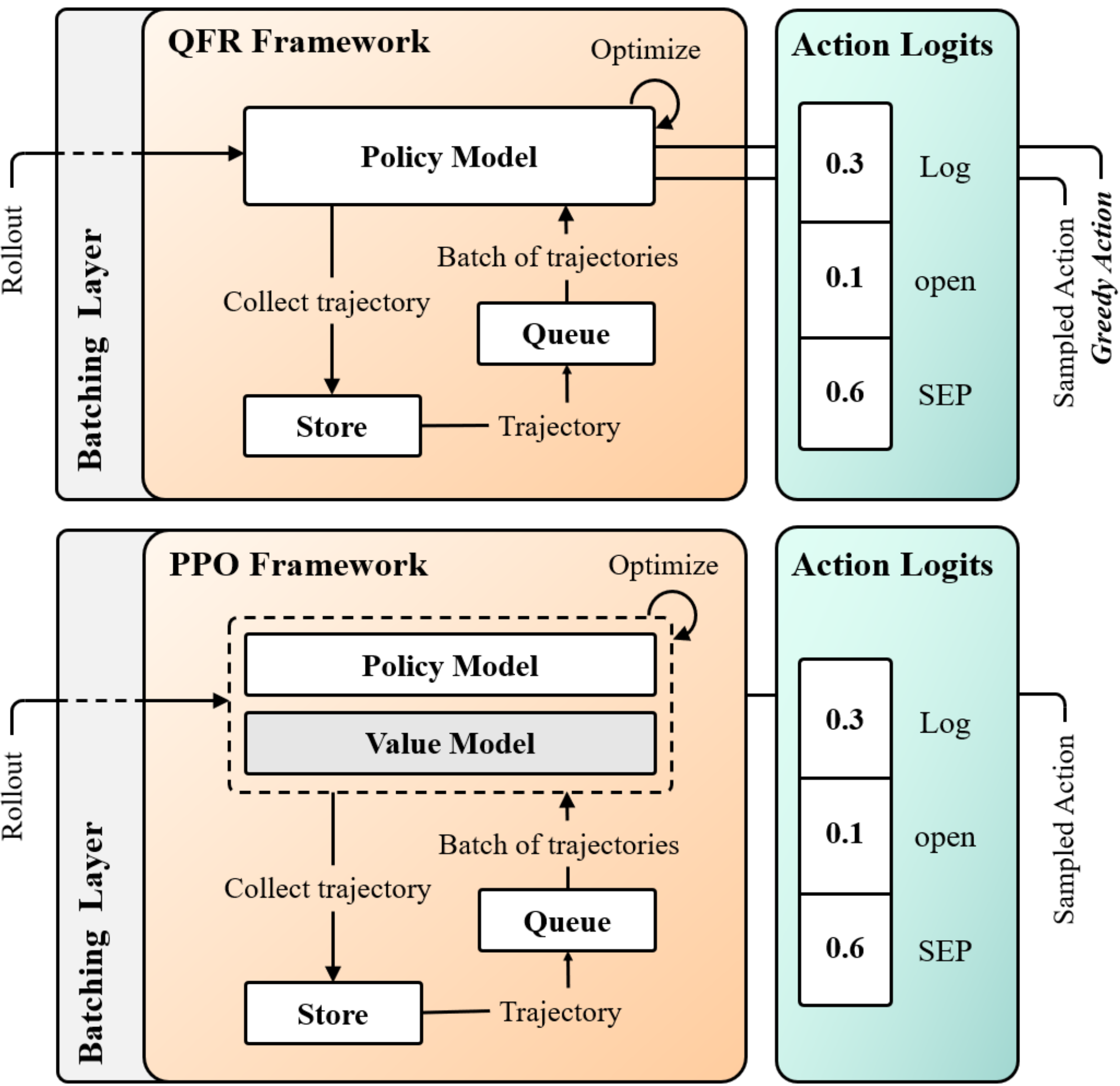}
\caption{Comparison between QFR and PPO. By discarding the critic network (value model), QFR requires much fewer rollouts than PPO, which significantly improves the overall convergence speed.}
\label{Figure2}
\end{figure}

\subsection{The Proposed Algorithm}
\label{Proposed Algorithm}
Inspired by the REINFORCE algorithm with baseline\cite{li2023remax}, the proposed QFR optimizes the gradient estimation by introducing a specific baseline value to reduce the variance:
\begin{equation}
    \widetilde{g}(\theta) = \frac{1}{N} \sum_{i=1}^{N} \sum_{t=1}^{T} \left[s_{\mathbf{\theta}}\left(\mathbf{a}_{1: t}^{i}\right) \left(r\left(\mathbf{a}_{1: T}^{i}\right) - r\left(\mathbf{\bar{a}}_{1: T}^i\right)\right)\right],    
    \label{QFR Loss 1}
\end{equation}

\noindent where the baseline value $r\left(\mathbf{\bar{a}}_{1: T}^i\right)$ can be obtained by greedily sampling a formulaic alpha factor with the maximal conditional probability and calculating the associated reward, i.e., \(
\bar{a}_{t} \in \operatorname{argmax} \pi_{\mathbf{\theta}}\left(\cdot \mid \mathbf{\bar{a}}_{1: t-1}\right)
\). 
The proposed baseline serves as a normalizing function by comparing the reward of the random response with that of the greedy response, thus can reduce the variance in the gradient estimation. While PPO can also reduce large variance during training, it does so at the cost of training an additional value model with similar complexity to the policy model. This not only slows down the convergence, but substantially increases the sampling time costs, as a forward pass must be performed for each state sampled from the buffer to obtain $V^\pi(\mathbf{s}_t)$. Comparatively, QFR receives the rewards as a trajectory feedback, which fits the nature of the underlying MDP, and saves much computational costs for rollout. The differences between QFR and PPO based formulaic alpha factors mining frameworks are depicted in Figure. \ref{Figure2}

\begin{figure*}[!h]
\centering
\includegraphics[width=0.8\textwidth]{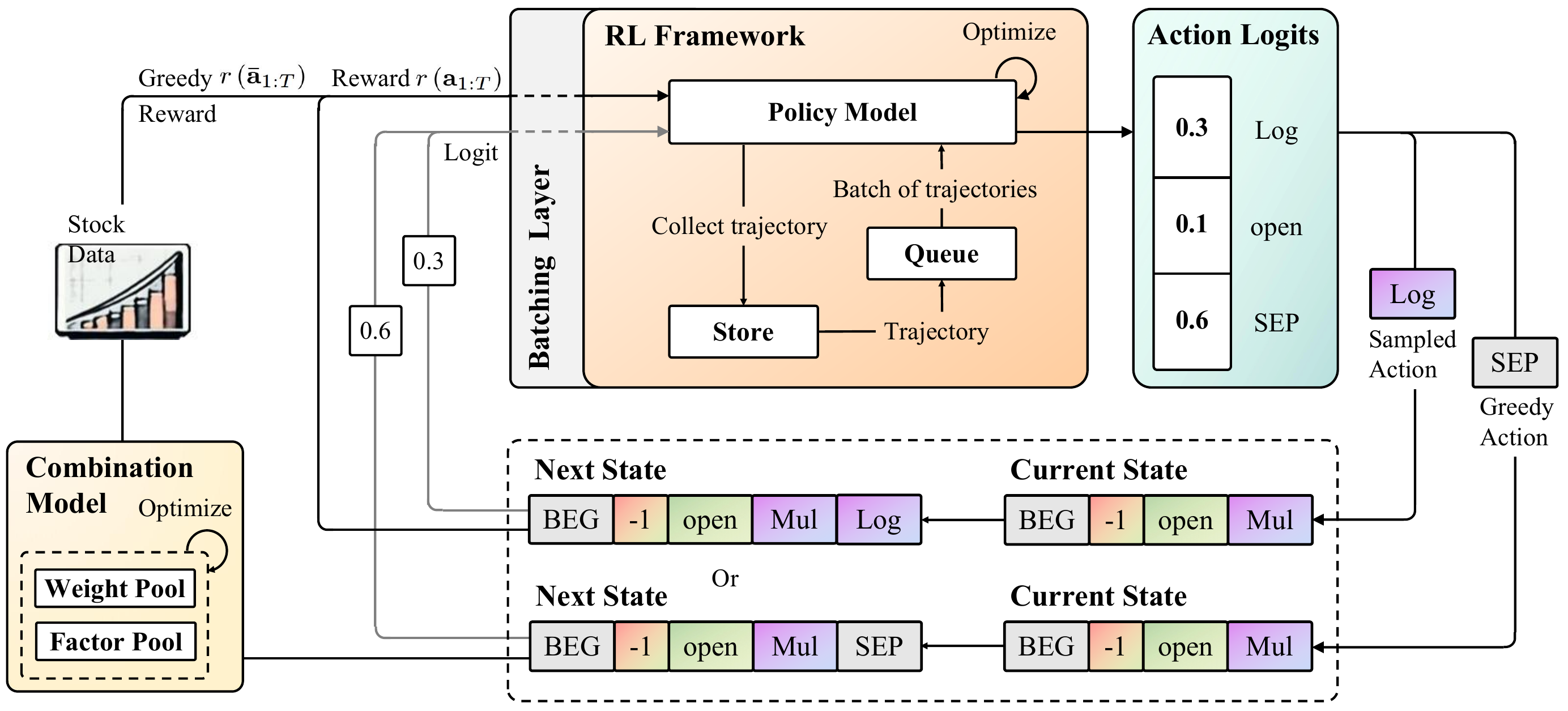}
\caption{The detailed pipeline for the proposed QFR algorithm. }
\label{Figure3}
\end{figure*}

The detailed pipeline of QFR is shown in Fig. \ref{Figure3}. The policy model generates one new token (action) at each step, and the sequence of generated tokens (states) can be uniquely converted into an expression based on RPN. The policy model uses random sampling and greedy sampling to complete two trajectories and generate two expressions, i.e., two factors. The combination model is used to maintain a weighted combination of principal factors and, at the same time, to evaluate these two factors. Specifically, the combined factor values are used to calculate \( r\left(\mathbf{a}_{1: T}\right) \) and \( r\left(\mathbf{\bar{a}}_{1: T}\right) \) along with the real asset data features $\mathcal{X}$. These two rewards, calculated by the proposed shaped reward function (detailed in Section \ref{rewardshaping}), form the estimation of (\ref{QFR Loss 1}). The algorithmic steps are given in Algorithm 1. 

\begin{algorithm}[b]  
	\caption{QFR}
	\LinesNumbered 
	\KwIn{Real asset price dataset $\mathcal{Y}$, Real asset feature dataset $\mathcal{X}=\{\mathbf{X}^{\prime}_l\}$, Initial policy weight $\mathbf{\theta}$, Initial combination model weight $\mathbf{\omega}$, Reward balancing coefficient $\lambda$.}
	\KwOut{Formulaic alpha factors generator $\pi_\mathbf{\theta}$.}
	\While{not converged}{
        Construct a normal factor $f_n$ with $\pi_{\mathbf{\theta}}\left(\cdot \mid \mathbf{a}_{1: t-1}\right)$\;
        Construct a greedy factor $f_g$ with $\pi_{\mathbf{\theta}}\left(\cdot \mid \mathbf{\bar{a}}_{1: t-1}\right)$\;
        Compute normal factor values $\{\mathbf{z}_{n,l}\}=\{f_n(\mathbf{X}_l)\}$\;
        Compute greedy factor values $\{\mathbf{z}_{g,l}\}=\{f_g(\mathbf{X}_l)\}$\;
        Compute $\{\mathbf{z}^{\prime}_{n,l}\}$ and $\{\mathbf{z}^{\prime}_{g,l}\}$ of the normal factor and greedy factor with $\mathbf{\omega}$, respectively\;
        Compute $r\left(\mathbf{a}_{1: T}\right)$ and $r\left(\mathbf{\bar{a}}_{1: T}\right)$ with the shaped reward and both $\{\mathbf{z}^{\prime}_{n,l}\}$ and $\{\mathbf{z}^{\prime}_{g,l}\}$ via Eq. (\ref{Shaped Reward})\;
		Update $\mathbf{\theta}$ via Eq. (\ref{QFR Loss 1})\;
        Update $\mathbf{\omega}$ via Eq. (\ref{combination model loss})\;	
	}
\label{ag1}
\end{algorithm}

To our best knowledge, we are the first to introduce this baseline in MDPs for formulaic alpha factors mining. Notably, while this baseline looks straightforward, it works very well for the underlying MDP. Moreover, we provide comprehensive and detailed theoretical understandings of why this baseline is particularly well suited to these MDPs in Section \ref{The Theoretical Analysis}.
\subsection{Time-varying Reward Shaping}
\label{rewardshaping}
The reward function in \cite{yu2023generating} focuses on the absolute excess returns of factors, while ignoring the risk-adjusted characteristics of the returns, i.e., how the factors are resistant to market volatility. In order to better balance returns and risks, QFR not only evaluate the predictive accuracy of factors, but considers the stability of their predictive signals, thus providing a more comprehensive assessment of factor performance.

Specifically, we introduce the Information Ratio (IR) for reward shaping. IR is a financial metric that measures the excess returns of factors relative to excess risk and is commonly used to evaluate the risk-adjusted performance of the factors:

\begin{equation}
    IR = \frac{\mathbb{E}_{t}\left[IC\left(\mathbf{z}_t, \mathbf{y}_t\right)\right]}{
\sqrt{\operatorname{Var}(IC\left(\mathbf{z}_t, \mathbf{y}_t\right))}}.
\label{Single IR}
\end{equation}

Equation (\ref{Single IR}) describes how the IR of a single factor is calculated. After the combination model, the IR for the factors becomes:
\begin{equation}
    \overline{IR} = \frac{\mathbb{E}_{t}\left[IC\left(\mathbf{z}^\prime_t, \mathbf{y}_t\right)\right]}{
\sqrt{\operatorname{Var}(IC\left(\mathbf{z}^\prime_t, \mathbf{y}_t\right))}}.
\end{equation}
Based on IR, we can construct a time-varying reward shaping mechanism. At the beginning of the training, a high tolerance is given to factors with low IR. As the training progresses and the quality of the generated factors improves, negative rewards are assigned to factors with low IR:
\begin{equation}
   r\left(\mathbf{a}_{1: T}\right)=\overline{IC} - \lambda \mathbb{I} \{\overline{IR}\leq \operatorname{clip}\left[ (t-\alpha) \cdot \eta, 0, \delta \right] \}.
   \label{Shaped Reward}
\end{equation}

\begin{figure}[t]
\centering
\includegraphics[width=0.43\textwidth]{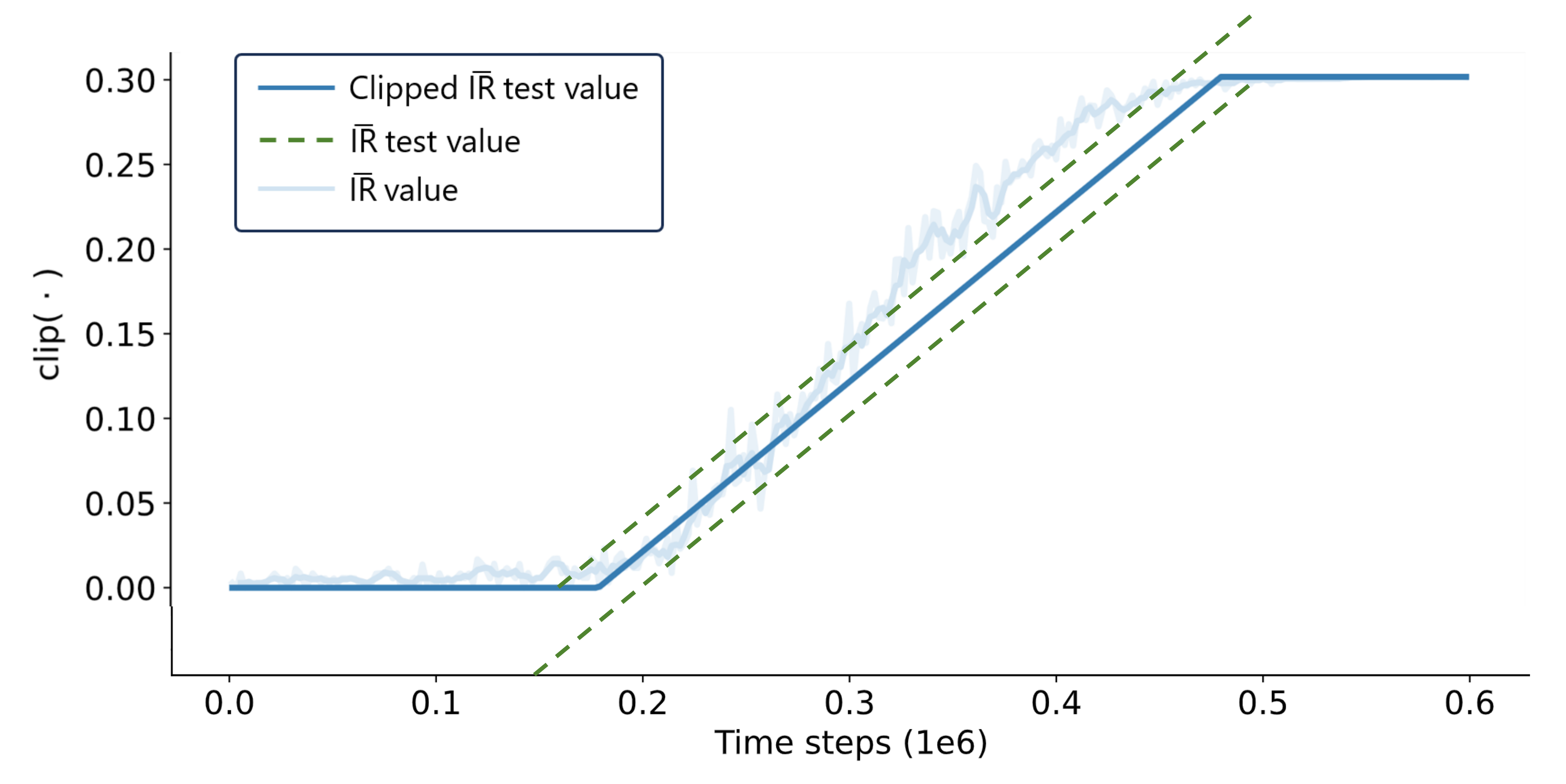}
\caption{Clip function values versus training time steps, with an example of $\overline{IR}$ values. When the clipped function value, i.e., the $\overline{IR}$ test value, is greater than the $\overline{IR}$ value of the generated combined factor, the factor quality is assumed to be poor, and the last step of the trajectory will receive a discounted $\overline{IC}$ as a reward. Conversely, a complete $\overline{IC}$ is obtained as a reward.}
\label{Figure4}
\end{figure}

Equation (\ref{Shaped Reward}) is the proposed shaped reward. $\mathbb{I}$ denotes the indicator function, which takes the value of 1 if the condition inside the parentheses is satisfied, and 0 otherwise. The clip function is defined as $\text{clip}(x, a, b) = \min\left(\max(x, a), b\right)\notag$, which is utilized to constrain the range of values. $\alpha$ is the time delay, $\eta$ is the slope of the change, $\delta$ is the maximum IR test value, and $\lambda$ is the shaping coefficient. The varying of the clip function over training time is illustrated in Fig. \ref{Figure4}. 

Using the IR test as reward shaping can force QFR focusing more on the long-term stability and consistency of the factors during optimization, rather than just the short-term profit peaks.
\subsection{The Theoretical Analysis}
\label{The Theoretical Analysis}
We provide a set of theoretical results for Algorithm 1, which includes the analysis of the training variance under different state transition functions in \textbf{Proposition 2}, the derivation of the upper bound of the variance in \textbf{Proposition 3}, and the demonstration that the variance decreases relative to the REINFORCE algorithm in \textbf{Proposition 4}. To justify the algorithm design, we first prove that the variance of the proposed algorithm is bounded.

\textbf{Proposition 1.} \textit{The gradient estimator (\ref{QFR Loss 1}) is unbiased for the objective function (\ref{original objective function}), i.e., $\mathbb{E}[\widetilde{g}(\mathbf{\theta})]=$ $\nabla_\mathbf{\theta} \mathbb{E}_{\mathbf{a}_{1: t} \sim \pi_\mathbf{\theta}}\left[r\left(\mathbf{a}_{1: T}\right)\right]$. }

\begin{proof}
We take the expectation over the randomness of responses $\left(\mathbf{a}_{1: t}^1, \ldots, \mathbf{a}_{1: t}^N\right)$ :
\begin{align}
\mathbb{E}[\widetilde{g}(\theta)] & = \mathbb{E}\Bigg[ \sum_{t=1}^T s_{\mathbf{\theta}}\left(\mathbf{a}_{1: t}\right) \times\left(r\left(\mathbf{a}_{1: T}\right)-r\left(\overline{\mathbf{a}}_{1: T}\right)\right)\Bigg] \nonumber\\
& = \mathbb{E}\left[\sum_{t=1}^T \nabla_\mathbf{\theta} \log \pi_\mathbf{\theta}\left(a_t \mid \mathbf{a}_{1: t-1}\right) r\left(\mathbf{a}_{1: T}\right)\right] \nonumber \nonumber\\
& \quad - \mathbb{E}\left[\sum_{t=1}^T \nabla_\mathbf{\theta} \log \pi_\mathbf{\theta}\left(a_t \mid \mathbf{a}_{1: t-1}\right) r\left( \mathbf{\bar{a}}_{1: T}\right)\right] \nonumber\\
& =\nabla_\mathbf{\theta} \mathbb{E}_{\mathbf{a}_{1: t} \sim \pi_\mathbf{\theta}}\left[ r\left(\mathbf{a}_{1: T}\right)\right] \nonumber \\
& \quad -\mathbb{E}_{\mathbf{a}_{1: t} \sim \pi_\theta}\left[\nabla_\theta \sum_{t=1}^T \log \pi_\theta\left(a_t \mid \mathbf{a}_{1: t-1}\right) \times r\left(\mathbf{\bar{a}}_{1: T}\right)\right] \label{log-derivative trick} \\
& = \nabla_\mathbf{\theta} \mathbb{E}_{\mathbf{a}_{1: t} \sim \pi_\mathbf{\theta}}\left[ r\left(\mathbf{a}_{1: T}\right)\right], \label{Bartlett identity}
\end{align}

where (\ref{log-derivative trick}) follows the so-called log-derivative trick, (\ref{Bartlett identity}) follows the idea of Bartlett identity $\sum_z \nabla_\mathbf{\theta} p_\mathbf{\theta}(z) b=\nabla_\mathbf{\theta}\left[\sum_z p_\mathbf{\theta}(z) b\right]=\nabla_\mathbf{\theta}[1 \cdot b]=0$, in which $p_\mathbf{\theta}$ can be any distribution and $b$ is constant\cite{bartlett1953approximate}. Notice that $\mathbf{\bar{a}}_{1: T}$ is conditionally independent on $\mathbf{\theta}$, due to the greedy sampling and apply $p_\mathbf{\theta}$ to the the distribution $\pi_\mathbf{\theta}\left(\bar{a}_{1: T} \mid x\right)$. Then, regarding the second item, we only need to consider \( \mathbb{E}_{\mathbf{a}_{1: T} \sim \pi_\theta}\left[\nabla_\theta \sum_{t=1}^T \log \pi_\theta(a_t \mid \mathbf{a}_{1: t-1})\right] = \nabla_\theta \mathbb{E}_{\mathbf{a}_{1: T} \sim \pi_\theta}\left[\sum_{t=1}^T \log \pi_\theta(a_t \mid \mathbf{a}_{1: t-1})\right] \), which is the sum of log probabilities, whose expected value (i.e., the average log probability) with respect to \( \mathbf{\theta} \) should be equal to 0, because the probability distribution \( \pi_\mathbf{\theta} \) is fixed under its own expectation. The proof is thus completed.
\end{proof}

Since the baseline value introduced by QFR is determined by the reward of the greedy policy, and corresponds to the reward distribution, it is statistically independent of the samples sampled by the normal policy, satisfying the requirements for the proof of unbiasedness.

\textbf{Proposition 2.} \textit{Given the deterministic transition function $\mathcal{T}(\mathbf{s}_{t+1}^\prime|\mathbf{s}_t,a_t)$, which satisfies the Dirac distribution, and the probabilistic transition function $\mathcal{T}(\mathbf{s}_{t+1}^{\prime\prime}|\mathbf{s}_t,a_t)$, if they are unbiased, we have \(\operatorname{Var}\left[\mathbf{s}_{T}^{D}\right] \approx \operatorname{Var}\left[\mathbf{s}_{T}^{R}\right] - \sum_{t = 1}^{T-1} \mathbb{E}_{\mathbf{s}_{t}^{\prime R}}\left[\operatorname{Var}\left[\mathbf{s}_{t}^{\prime \prime R}|\mathbf{s}_{t}^{\prime R}\right]\right]\), i.e., \(\operatorname{Var}\left[\mathbf{s}_{T}^{D}\right] \leq \operatorname{Var}\left[\mathbf{s}_{T}^{R}\right]\), where $\mathbf{s}_{T}^{D}$ denotes a possible sequence generated by $\mathcal{T}(\mathbf{s}_{t+1}^\prime|\mathbf{s}_t,a_t)$, and $\mathbf{s}_{T}^{R}$ is a sequence from $\mathcal{T}(\mathbf{s}_{t+1}^{\prime\prime}|\mathbf{s}_t,a_t)$. }

\begin{proof}
    The transition model of MDPs defined in Section \ref{Alpha Factor Mining Processes} can be decomposed as follows. Firstly, the policy model gives a token $\mathbf{a}_{t}$ based on the input sequence $\mathbf{s}_{t} = \mathbf{a}_{1:t-1}$, and the output sequence $\mathbf{s}_{t+1}^\prime$ is built by simply appending $\mathbf{a}_{t}$ to $\mathbf{s}_{t}$, i.e., $\mathbf{s}_{t+1}^\prime = \mathbf{a}_{1:t}$. If the transition is deterministic, meaning the transition function satisfies the Dirac distribution, then $\mathbf{s}'_{t+1}$ is directly used as the input state for next step. If a probabilistic transition function is used, $\mathcal{T}(\mathbf{s}''_{t+1}|\mathbf{s}_{t},a_t)$ is applied and it generates a new random sequence $\mathbf{s}''_{t+1}$ as the next input state.

Firstly, we estimate the variance of sequences during the appending process (\,$\mathbf{s}_{t} \xrightarrow{\pi (a_t|\mathbf{s}_{t})}\mathbf{s}'_{t+1}$\,). Given that $\mathbf{s}_{t} = \mathbf{a}_{1:t-1}$ is a vector of length $t-1$, its variance can be expanded as the summation of its all digits:
\begin{align}
    \operatorname{Var}\left[\mathbf{s}_{t}\right]& = \mathbb{E}\left[ \left(\mathbf{s}_{t} - \mathbb{E} \left[\mathbf{s}_{t}\right] \right)^2\right]
    = \sum_{i=1}^{t-1}\mathbb{E}\left[ \left(a_i-\mathbb{E}\left[a_{i}\right]\right)^2\right]\nonumber\\
    &=\sum_{i=1}^{t-1} \operatorname{Var} \left[ a_i\right] \nonumber.
\end{align}

Similarly, the variance of $\mathbf{s}'_{t+1}$ is:
\begin{align}
    \operatorname{Var}\left[\mathbf{s}'_{t+1}\right]& = \sum_{i=1}^{t} \operatorname{Var}  \left[a_i\right] \nonumber\\
    & = \operatorname{Var}\left[\mathbf{s}_{t}\right] + \mathbb{E}_{\mathbf{s}_{t}}\left[ \operatorname{Var} \left[a_t|\mathbf{s}_{t}\right]\right].\label{variance1}
\end{align}

The equation holds because $\mathbf{s}'_{t+1}$ and $\mathbf{s}_{t}$ has the same first $t-1$ digits, while $\mathbf{s}'_{t+1}$ has one more token at the  $t$-th position, which naturally introduces more variance. The variance originates from the policy model and can be expressed as
$\mathbb{E}_{\mathbf{s}_{t}}\left[ \operatorname{Var} \left[a_t|\mathbf{s}_{t}\right]\right] = \sum_{a_t, \mathbf{s}_{t}}P(\mathbf{s}_{t})\pi\left(a_t|\mathbf{s}_{t}\right)\left(a_t - \mathbb{E}\left[{a_t|\mathbf{s}_{t}}\right]\right)^2$

Then, we estimate the variance of sequences after the transition process (\,$\mathbf{s}'_{t+1}\xrightarrow{\mathcal{T}(\mathbf{s}''_{t+1}|\mathbf{s}_{t},a_t)}\mathbf{s}''_{t+1}$\,). Since the information of $\mathbf{s}_{t}$ and $a_{t}$ are completely included by $\mathbf{s}'_{t+1}$, the transition function $\mathcal{T}(\mathbf{s}''_{t+1}|\mathbf{s}_{t},a_t)$ can be written as $\mathcal{T}(\mathbf{s}''_{t+1}|\mathbf{s}'_{t+1})$ and is then applied on $\mathbf{s}'_{t+1}$. The variance of $\mathbf{s}''_{t+1}$ can be calculated by:
\begin{align}
    \operatorname{Var} \left[\mathbf{s}''_{t+1}\right]
    & = \mathbb{E}_{\mathbf{s}'_{t+1}}\left[\mathbb{E}_{\mathbf{s}''_{t+1}}\left[(\mathbf{s}''_{t+1})^2 | \mathbf{s}'_{t+1} \right]\right]- (\mathbb{E}\left[\mathbf{s}''_{t+1}\right])^2 \nonumber\\
    & = \mathbb{E}_{\mathbf{s}'_{t+1}}\left[\operatorname{Var}\left[\mathbf{s}''_{t+1}|\mathbf{s}'_{t+1}\right] + (\mathbb{E}\left[\mathbf{s}''_{t+1}|\mathbf{s}'_{t+1}\right])^2 \right]\nonumber\\
    & \quad- (\mathbb{E}\left[\mathbf{s}''_{t+1}\right])^2.\label{variance2}
\end{align}

Assuming the transition model is unbiased, we have $\mathbb{E}\left[\mathbf{s}''_{t+1}|\mathbf{s}'_{t+1}\right] = \mathbf{s}'_{t+1}$ and $\mathbb{E}\left[\mathbf{s}''_{t+1}\right] = \mathbb{E}\left[\mathbf{s}'_{t+1}\right]$. Then (\ref{variance2}) can be further simplified:
\begin{align}
\operatorname{Var} \left[\mathbf{s}''_{t+1}\right]& =\mathbb{E}_{\mathbf{s}'_{t+1}}\left[\operatorname{Var}\left[\mathbf{s}''_{t+1}|\mathbf{s}'_{t+1}\right] + (\mathbf{s}'_{t+1})^2 \right]\nonumber\\ 
&\quad - (\mathbb{E}\left[\mathbf{s}'_{t+1}\right])^2 \nonumber\\
&=
\mathbb{E}_{\mathbf{s}'_{t+1}}\left[\operatorname{Var}\left[\mathbf{s}''_{n}|\mathbf{s}'_{t+1}\right]\right] + \mathbb{E}\left[(\mathbf{s}'_{t+1})^2\right]\nonumber\\
& \quad - (\mathbb{E}\left[\mathbf{s}'_{t+1}\right])^2 \nonumber\\
&=
\mathbb{E}_{\mathbf{s}'_{t+1}}\left[\operatorname{Var}\left[\mathbf{s}''_{t+1}|\mathbf{s}'_{t+1}\right]\right] + \operatorname{Var} \left[\mathbf{s}'_{t+1}\right],\label{variance3}
\end{align}

where \(\mathbb{E}_{\mathbf{s}_{t+1}^{\prime}}\left[\operatorname{Var}\left[\mathbf{s}_{t+1}^{\prime \prime}|\mathbf{s}_{t+1}^{\prime}\right]\right]\) can be expressed as:
\begin{align}
    &\mathbb{E}_{\mathbf{s}_{t+1}^{\prime}}\left[\operatorname{Var}\left[\mathbf{s}_{t+1}^{\prime \prime}|\mathbf{s}_{t+1}^{\prime}\right]\right] \nonumber\\
    = &\sum_{\mathbf{s}_{t+1}^{\prime}, \mathbf{s}_{t+1}^{\prime  \prime}}P(\mathbf{s}_{t+1}^{\prime})T\left(\mathbf{s}_{t+1}^{\prime  \prime}|\mathbf{s}_{t+1}^{\prime}\right)\left(\mathbf{s}_{t+1}^{\prime  \prime} - \mathbf{s}_{t+1}^{\prime}\right)^2.\nonumber
\end{align}

Finally, we can estimate the variance under deterministic transition by recursively using (\ref{variance1}):
\begin{align}
\operatorname{Var}\left[\mathbf{s}_{T}^{D}\right]
    & = \sum_{t = 1}^{T-1} \mathbb{E}_{\mathbf{s}_{t}^{D}}\left[ \operatorname{Var} \left[a_t|\mathbf{s}_{t}^{D}\right]\right]\nonumber.
\end{align}

The variance under probabilistic transition can be estimated using (\ref{variance1}), (\ref{variance3}):
\begin{align}
&\operatorname{Var}\left[\mathbf{s}_{T}^{R}\right]\nonumber
    \\=&\sum_{t = 1}^{T-1} \mathbb{E}_{\mathbf{s}_{t}^{R}}\left[ \operatorname{Var} \left[a_t|\mathbf{s}_{t}^{R}\right]\right] +  \sum_{t = 1}^{T-1} \mathbb{E}_{\mathbf{s}_{t+1}^{\prime R}}\left[\operatorname{Var}\left[\mathbf{s}_{t+1}^{\prime \prime R}|\mathbf{s}_{t+1}^{\prime R}\right]\right]\nonumber.
\end{align}

Since $\mathbb{E}_{\mathbf{s}_{t}^{D}}\left[ \operatorname{Var} \left[a_t|\mathbf{s}_{t}^{D}\right]\right]$ and $\mathbb{E}_{\mathbf{s}_{t}^{D}}\left[ \operatorname{Var} \left[a_t|\mathbf{s}_{t}^{D}\right]\right]$ both have the order of $\sim\text{max}(|a_t|^2)$, we assume $\mathbb{E}_{\mathbf{s}_{t}^{D}}\left[ \operatorname{Var} \left[a_t|\mathbf{s}_{t}^{D}\right]\right] \approx \mathbb{E}_{\mathbf{s}_{t}^{R}}\left[ \operatorname{Var} \left[a_t|\mathbf{s}_{t}^{R}\right]\right]$. Finally the variances under both cases approximately satisfy the following condition:
\begin{align}
\operatorname{Var}\left[\mathbf{s}_{T}^{D}\right] \approx \operatorname{Var}\left[\mathbf{s}_{T}^{R}\right] - \sum_{t = 1}^{T-1} \mathbb{E}_{\mathbf{s}_{t+1}^{\prime R}}\left[\operatorname{Var}\left[\mathbf{s}_{t+1}^{\prime \prime R}|\mathbf{s}_{t+1}^{\prime R}\right]\right]. \nonumber
\end{align}

The equation shows that $\mathbf{s}_{T}^{D}$ has a lower variance than $\mathbf{s}_{T}^{R}$ by $\sum_{t = 1}^{T-1} \mathbb{E}_{\mathbf{s}_{t+1}^{\prime R}}\left[\operatorname{Var}\left[\mathbf{s}_{t+
1}^{\prime \prime R}|\mathbf{s}_{t+1}^{\prime R}\right]\right]$. This relation holds because the variance of $\mathbf{s}_{T}^{D}$ only originates from randomness of decision making during each step, while the variance of $\mathbf{s}_{T}^{R}$ is also from the random transition process. The proof is thus completed.
\end{proof}

This proposition shows that MDPs with deterministic state transition functions (following the Dirac distribution) exhibit lower variance compared to those with probabilistic state transition functions. Consequently, this helps mitigate the high variance issue inherent in REINFORCE.

\textbf{Proposition 3.} \textit{Consider the parameterization $\pi_\mathbf{\theta}(a \mid \mathbf{a}_{1: t-1})=\exp \left(\mathbf{\theta}_{a}\right) / \sum_{a^{\prime}} \exp \left(\mathbf{\theta}_{ a^{\prime}}\right)$, the variance of the gradient estimator (\ref{QFR Loss 1}) is bounded by $8 \times r_{\max }^2 \times T^2  / N$. Generally, (\ref{QFR Loss 1}) is bounded by $c \times r_{\max }^2 \times T^2 \times S^2 / N$, where $c$ is a universal constant, $S$ is an upper bound of $\left\|\nabla_\mathbf{\theta} \log \pi_\mathbf{\theta}\left(a_t \mid a_{1: t-1}\right)\right\|$ for all $\left(\mathbf{\theta}, \mathbf{a}_{1: t}\right)$, and $r_{\max }=\max _{\mathbf{a}_{1: t}}\left|r\left(\mathbf{a}_{1: T}\right)\right|$.}

\begin{proof}
We first define $\widetilde{g}_i(\mathbf{\theta})$ as the gradient calculated on the i-th sample $\left(\mathbf{a}_{1: T}^i\right)$:
\begin{equation*}
    \widetilde{g}_i(\mathbf{\theta})= \sum_{t=1}^T\left[s_\mathbf{\theta}\left(\mathbf{a}_{1: t}^i\right) \times\left(r\left( \mathbf{a}_{1: T}^i\right)-r\left(\mathbf{\bar{a}}_{1: T}^i\right)\right)\right].
\end{equation*}

Since different samples $\left(\mathbf{a}_{1: t}^i\right), \forall i \in[N]$ are independent, we have:
\begin{equation*}
    \operatorname{Var}[\widetilde{g}(\mathbf{\theta})]=\operatorname{Var}\left[\frac{1}{N} \sum_{i=1}^N \widetilde{g}_i(\mathbf{\theta})\right]=\frac{1}{N^2} \sum_{i=1}^N \operatorname{Var}\left[\widetilde{g}_i(\mathbf{\theta})\right].    
\end{equation*}

For each $i \in[N]$, we have:

\begin{align}
\operatorname{Var}\left[\widetilde{g}_i(\mathbf{\theta})\right] & =\mathbb{E}\left[\left\|\widetilde{g}_i(\mathbf{\theta})\right\|^2\right]-\left\|\mathbb{E}\left[\widetilde{g}_i(\mathbf{\theta})\right]\right\|^2 \nonumber\\
& \leq \mathbb{E}\left[\left\|\widetilde{g}_i(\mathbf{\theta})\right\|^2\right] \nonumber\\
& =\mathbb{E}\left[\left(r\left(\mathbf{a}_{1: T}^i\right)-r\left(\mathbf{\bar{a}}_{1: T}^i\right)\right)^2\left\|\sum_{t=1}^T s_\mathbf{\theta}\left(\mathbf{a}_{1: t}^i\right)\right\|^2\right] \nonumber\\
& \leq 4 r_{\max }^2 \mathbb{E}\left[\left\|\sum_{t=1}^T s_\mathbf{\theta}\left(\mathbf{a}_{1: t}^i\right)\right\|^2\right] \label{inequality property}\\
& \leq 4 r_{\max }^2 T^2\left(\max _{\mathbf{a}_{1: t}}\left\|s_\mathbf{\theta}\left(\mathbf{a}_{1: t}\right)\right\|\right)^2 \label{triangle inequality}\\
& \leq 4 r_{\max }^2 T^2 S^2 \nonumber.
\end{align}

(\ref{inequality property}) follows from the property of inequalities, specifically \((r(a) - r(b))^2 \leq 2(r(a)^2 + r(b)^2) \leq 4r_{\max}^2\), where \(r_{\max}\) is the maximum value of the reward function. (\ref{triangle inequality}) follows from the triangle inequality. Then we have:
\begin{equation*}
    \operatorname{Var}[\widetilde{g}(\mathbf{\theta})]=\frac{1}{N^2} \sum_{i=1}^N \operatorname{Var}\left[\widetilde{g}_i(\mathbf{\theta})\right] \leq \frac{4 r_{\max }^2 T^2 S^2}{N} .
\end{equation*}

Specially, the upper bound for $S$ can be estimated if we take derivatives only with respect to variables before the softmax function. i.e, to take derivatives with respect to $\theta_a$ where $\pi_\theta(a \mid \mathbf{a}_{1: t-1})=\exp \left(\theta_{a}\right) / \sum_{a^{\prime}} \exp \left(\theta_{ a^{\prime}}\right)$:
\begin{align}
&\quad \frac{\partial }{\partial \theta_{a'}}\log \pi_\theta\left(a \mid \mathbf{a}_{1: t-1}\right)\notag \\
& =\frac{\partial }{\partial \theta_{a'}} \left(\theta_a - \log \left( \sum_{a^{\prime}} \exp \left(\theta_{ a^{\prime}}\right)\right) \right)\notag\\
& = \delta_{a,a'} - \exp \left(\theta_{a}\right) / \sum_{a^{\prime}} \exp \left(\theta_{ a^{\prime}}\right)\notag\\
&=\left\{\begin{array}{ll}  1 - \pi_\theta(a \mid \mathbf{a}_{1: t-1}) & \text { if } a' = a \\ \pi_\theta(a' \mid \mathbf{a}_{1: t-1}) & \text { otherwise, }\end{array}\right. \nonumber
\end{align}

Consequently the upper bound for $S$ is:
\begin{align}
S &= \max \left\|\nabla_\mathbf{\theta} \log \pi_\mathbf{\theta}\left(a_t \mid a_{1: t-1}\right)\right\| \notag\\
&=\max \sqrt{\left(1 - \pi_\theta(a \mid \mathbf{a}_{1: t-1})\right)^2 + \sum_{a'\neq a} \pi_\theta(a' \mid \mathbf{a}_{1: t-1})^2}\notag\\
&\leq \sqrt{1+1} = \sqrt{2}, \nonumber
\end{align}

then we have:
\begin{equation*}
\operatorname{Var}[\widetilde{g}(\mathbf{\theta})]\leq \frac{8 r_{\max }^2 T^2}{N}.    
\end{equation*}

The proof is thus completed. Notably, the noise in real asset data features $\mathcal{X}$ potentially increases \( r_{\text{max}} \). This could result in an increase in the upper bound of \( 8r_{\text{max}}^2T^2/N \), but the proposition remains correct. 
\end{proof}

This proposition demonstrates that the variance of the QFR algorithm is bounded. Specifically, the proof process defines the gradient calculation formula based on the i-th sample and uses properties of inequalities and the triangle inequality to derive an upper bound for the variance of the gradient for each sample, which implies that the stability of the algorithm during the estimation process is guaranteed.

\textbf{Proposition 4.} \textit{For any 2-armed bandit, consider the parameterization $\pi_\mathbf{\theta}(a \mid \mathbf{a}_{1: t-1})=\exp \left(\mathbf{\theta}_{a}\right) / \sum_{a^{\prime}} \exp \left(\mathbf{\theta}_{ a^{\prime}}\right)$, where $\mathbf{\theta} \in \mathbb{R}^{|\mathcal{X}| \times 2}$ with $|\mathcal{X}|$ being the context size and 2 being the action size. Assume $a_1$ is the optimal action and rewards are positive. Then, if $\pi_\mathbb{\theta}\left(a_1 \mid \mathbf{a}_{1: t-1}\right) \leq 0.5+0.5 r_2 /\left(r_1-r_2\right)$, we have
\(
\begin{gathered}
\operatorname{Var}[\widetilde{g}(\theta)]<\operatorname{Var}[\widehat{g}(\theta)] \text {.} 
\end{gathered}
\)
Notably, if $r_1 < 2 r_2$, any possible $\pi_\mathbb{\theta}\left(a_1 \mid \mathbf{a}_{1: t-1}\right) \in (0,1)$ guarantees a lower variance using QFR.}

\begin{proof}
    Firstly, we have the definition of two kinds of policy gradient $\hat{g}(\mathbf{\theta})=\sum_{t=1}^T \nabla_\mathbf{\theta} \ln \pi_\mathbf{\theta}\left(a_t \mid \mathbf{a}_{1: t-1}\right) r\left(\mathbf{a}_{1: T}\right)$ and $\tilde{g}(\mathbf{\theta})=\sum_{t=1}^T \nabla_\mathbf{\theta} \ln \pi_\mathbf{\theta}\left(a_t \mid \mathbf{a}_{1: t-1}\right)(r\left(\mathbf{a}_{1: T}\right)-r\left(\overline{\mathbf{a}}_{1: T}\right))$. We want to prove that the variance of QFR decreases relative to the REINFORCE algorithm:
\begin{equation}
\operatorname{Var}_{s,  a}\left[\tilde{g}(\mathbf{\theta})\right]<\operatorname{Var}_{s,  a }\left[\hat{g}(\mathbf{\theta})\right],\nonumber
\end{equation}

or a stronger result:
\begin{equation}
\operatorname{Var}_{ a  }\left[\tilde{g}(\mathbf{\theta})\right]<\operatorname{Var}_{ a }\left[\hat{g}(\mathbf{\theta})\right].\nonumber
\end{equation}

For simplicity, we consider the gradient $g(\mathbf{\theta})$ only contains one sample at time $t$. Proposition 1 proves that they have same expectation value, Then we have:

\begin{align}
 & \quad\operatorname{Var}\left[\tilde{g}(\mathbf{\theta})\right]-\operatorname{Var}\left[\hat{g}(\mathbf{\theta})\right] \nonumber\\
& = \mathbb{E}_{\mathbf{a}_{1: t}\sim \pi_\mathbf{\theta}} \left[\tilde{g}(\mathbf{\theta})^2\right]-\left(\mathbb{E}_{\mathbf{a}_{1: t}\sim \pi_\mathbf{\theta}}[\tilde{g}(\mathbf{\theta})]\right)^2\nonumber \\
& - \mathbb{E}_{\mathbf{a}_{1: t}\sim \pi_\mathbf{\theta}}\left[\hat{g}(\mathbf{\theta})^2\right] + \left(\mathbb{E}_{\mathbf{a}_{1: t}\sim \pi_\mathbf{\theta}}[\hat{g}(\mathbf{\theta})]\right)^2 \label{vartoexpt}\\
& = \mathbb{E}_{\mathbf{a}_{1: t}\sim \pi_\mathbf{\theta}}\left[\tilde{g}(\mathbf{\theta})^2\right]-\mathbb{E}_{\mathbf{a}_{1: t}\sim \pi_\mathbf{\theta}}\left[\hat{g}(\mathbf{\theta})^2\right] \label{expcancel}\\
& = \mathbb{E}_{\mathbf{a}_{1: t} \sim \pi_\mathbf{\theta}}\left[\left(\nabla_\theta \ln \pi_\theta\left(a_t \mid \mathbf{a}_{1: t-1}\right)\right)^2\right.\nonumber\\
&\quad\times \left.\left((r\left(\mathbf{a}_{1: T}\right)-r\left(\overline{\mathbf{a}}_{1: T}\right))^2 -r\left(\mathbf{a}_{1: T}\right)^2\right)\right] \nonumber\\
& = \mathbb{E}_{\mathbf{a}_{1: t} \sim \pi_\mathbf{\theta}}\left[\left(\nabla_\theta \ln \pi_\theta\left(a_t \mid \mathbf{a}_{1: t-1}\right)\right)^2\right.\nonumber\\
&\quad\times\left.\left(-2 r(\mathbf{a}_{1: T}) r\left(\overline{\mathbf{a}}_{1: T}\right)+r\left(\overline{\mathbf{a}}_{1: T}\right)^2\right)\right] \nonumber\\
& = r\left(\overline{\mathbf{a}}_{1: T}\right)\mathbb{E}_{\mathbf{a}_{1: t} \sim \pi_\mathbf{\theta}}\left[\left(\nabla_\theta \ln \pi_\theta\left(a_t \mid \mathbf{a}_{1: t-1}\right)\right)^2\right.\nonumber\\ 
&\quad \times\left.(-2 r\left(\mathbf{a}_{1: T}\right)+r\left(\overline{\mathbf{a}}_{1: T}\right))\right],\label{getrout}
\end{align}

where (\ref{vartoexpt}) follows from the expanded form of variance, and in (\ref{expcancel}) two expectation terms canceled out, based on Proposition 1 that both gradient estimators have the same expectation value. (\ref{getrout}) holds because the greedy reward $r\left(\overline{\mathbf{a}}_{1: T}\right)$ is independent of action $a_t$.  

In (\ref{getrout}) the internal factor $(-2 r\left(\mathbf{a}_{1: T}\right)+r\left(\overline{\mathbf{a}}_{1: T}\right))$ tends to be negative and reduces the variance of QFR when the greedy reward is not too large. Consider a special case: the dimensionality of action space is 2. Let $p=\pi_1=\pi_\mathbb{\theta}\left(a_1 \mid \mathbf{a}_{1: t-1}\right)$ and $1-p=\pi_2=\pi_\mathbb{\theta}\left(a_2 \mid \mathbf{a}_{1: t-1}\right)$. Furthermore, let $a_1$ be the optimal action, $r_1>r_2$. By the parameterization $\pi_\theta(a \mid \mathbf{a}_{1: t-1})=\exp \left(\theta_{a}\right) / \sum_{a^{\prime}} \exp \left(\theta_{ a^{\prime}}\right)$, we have:

\begin{align}
\nabla_\theta \log \pi_\theta\left(a_1 \mid \mathbf{a}_{1: t-1}\right) & =\left(1-\pi_1,-\pi_2\right)^{\top} ,\nonumber\\
\nabla_\theta \log \pi_\theta\left(a_2 \mid \mathbf{a}_{1: t-1}\right) & =\left(1-\pi_2,-\pi_1\right)^{\top} .\nonumber
\end{align}

Under this special case, we can rewrite the difference between the variances:

\begin{align}
& \quad\operatorname{Var} [\tilde{g}(\mathbf{\theta})]-\operatorname{Var}[\hat{g}(\mathbf{\theta})] \nonumber\\
& = r\left(\overline{\mathbf{a}}_{1: T}\right)   \mathbb{E}_{\mathbf{a}_{1: t} \sim \pi_\mathbf{\theta}} \Big[\left(\nabla_\theta \ln \pi_\theta\left(a_t \mid \mathbf{a}_{1: t-1}\right)\right)^2\nonumber\\
&\quad\times(-2 r\left(\mathbf{a}_{1: T}\right)+r\left(\overline{\mathbf{a}}_{1: T}\right))\Big] \nonumber\\
& = r\left(\overline{\mathbf{a}}_{1: T}\right) \cdot\Big[2 p (1-p)^2 \left(-2 r_1+r\left(\overline{\mathbf{a}}_{1: T}\right)\right)\nonumber \\
&\quad + 2 (1-p) p^2 \left(-2 r_2+r\left(\overline{\mathbf{a}}_{1: T}\right)\right)\Big] \label{expand}\\
& = 2 p(1-p) r\left(\overline{\mathbf{a}}_{1: T}\right) \left(r\left(\overline{\mathbf{a}}_{1: T}\right)-2(1-p) r_1-2 p r_2\right), \nonumber
\end{align}
where (\ref{expand}) follows from enumerating the cases where the agent chooses between the two actions. Expanding the expectation, we can obtain: \(
\mathbb{E}_{\mathbf{a}_{1: t} \sim \pi_{\theta}}\left[\left(\nabla_{\theta} \ln \pi_{\theta}\left(a_{t} \mid \mathbf{a}_{1: t-1}\right)\right)^{2}\right]=\pi_{1}\left[\left(1-\pi_{1}\right)^{2}+{\pi_{2}}^{2}\right] + \pi_{2}\left[\left(1-\pi_{2}\right)^{2}+{\pi_{1}}^{2}\right] = 2 p (1-p)^2 + 2 (1-p) p^2
\).

For $p<1-p$ case, the sub-optimal action $a_2$ is dominated. The greedy policy prefers to select $a_2$, and thus $ r\left(\overline{\mathbf{a}}_{1: T}\right) = r_2$. Then the condition of a lower variance for QFR can be reduced to:
\begin{align}
&\quad\operatorname{Var} [\tilde{g}(\mathbf{\theta})]-\operatorname{Var}[\hat{g}(\mathbf{\theta})]\nonumber\\
&= 2 p(1-p) r_2\left(r_2-2(1-p) r_1-2 p r_2\right) < 0\nonumber\\
&\Longleftrightarrow p<1+\frac{r_2}{2\left(r_1-r_2\right)}\nonumber,
\end{align}
which is always satisfied, considering $p < 1/2$. 

For $p>1-p$ case, the optimal action $a_1$ is dominated, so $ r\left(\overline{\mathbf{a}}_{1: T}\right) = r_1$. We can similarly derive the condition of $p$:
\begin{align}
&\quad\operatorname{Var} [\tilde{g}(\mathbf{\theta})]-\operatorname{Var}[\hat{g}(\mathbf{\theta})]\nonumber \\
& = 2 p(1-p) r_1\left(r_1-2(1-p) r_1-2 p r_2\right) < 0\nonumber\\
&\Longleftrightarrow p<\frac{1}{2}+\frac{r_2}{2\left(r_1-r_2\right)} \nonumber.
\end{align}

Combining both cases, we come to the result that any $p$ that satisfies:
\begin{equation}
    0<p<\frac{1}{2}+\frac{r_2}{2\left(r_1-r_2\right)}, \nonumber
\end{equation}
will ensure $\operatorname{Var} [\tilde{g}(\mathbf{\theta})]<\operatorname{Var}[\hat{g}(\mathbf{\theta})]$. Notably, (\ref{getrout}) reveals that if $r_1 < 2 r_2$, any possible $p \in (0,1)$ guarantees a lower variance using QFR. This result extends naturally to larger action spaces: variance reduction occurs when $r\left(\mathbf{a}_{1: T}\right)>r\left(\overline{\mathbf{a}}_{1: T}\right)/2$. Given that industry benchmarks for high-quality factors rarely exceed ICs of $0.1$\cite{grinold2000active}, this condition holds even in the worst case once $r\left(\mathbf{a}_{1: T}\right)> 0.05$.
\end{proof}

Proposition 4 discloses that QFR achieves the variance reduction compared to REINFORCE when the optimal action has not dominated (e.g., $\left.\pi_\mathbb{\theta}\left(a_1 \mid \mathbf{a}_{1: t-1}\right) \leq 0.5\right)$.

\section{Numerical Results}
In this section, we numerically evaluate QFR, comparing it with both state-of-the-art RL algorithms and other factor mining methods. Our experimental study is comprised of six stock datasets (detailed in Section \ref{Environment Settings}), and evaluates the efficiency of various RL algorithms in solving MDPs for mining formulaic alpha factors in Section \ref{Comparisons with Other RLs}, after which we study the hyper-parameters of reward shaping in Section \ref{The Studies of the Time-varying Reward Shaping}, and considers five factor mining methods (discussed in Section \ref{FacEva} and \ref{Investment Simulation}). Finally, the ablation study confirms the importance of the two improvements in Section \ref{Ablation Study}.

\subsection{Environment Settings}
\label{Environment Settings}
The raw data sourced from the Chinese A-shares market, as well as the US stock market, specifically focusing on the constituent stocks of the China Securities Index 300 (CSI300, the index composed of the 300 most liquid and largest A-share stocks listed on the Shanghai and Shenzhen Stock Exchanges), the China Securities Index 500 (CSI500, the index representing 500 A-share stocks with mid-cap market values), the China Securities Index 1000 (CSI1000, an index including 1000 smaller-cap A-share stocks), the S\&P 500 Index (SPX, the Dow Jones Industrial Average, which tracks 30 major US blue-chip companies representing key sectors), the Dow Jones Industrial Average (DJI), and the NASDAQ 100 Index (NDX, the NASDAQ-100, consisting of 100 of the largest non-financial companies listed on the Nasdaq Stock Market) are utilized to model the MDPs for mining formulaic alphas in our experiment. Given many features including macroeconomic features, company's fundamental features, and stock price-volume features are not always publicly available, we have only identified six publicly available primary stock price-volume features for reproducibility to generate the formulaic alphas. They include the opening price (open), closing price (close), highest price (high), lowest price (low), trading volume (volume), and volume-weighted average price (vwap). Our objective is to generate formulaic alphas that exhibit a high IC with respect to the ground-truth 5-day asset returns. The dataset is split into three subsets: a training set spanning from 2016/01/01 to 2020/01/01, a validation set spanning from 2020/01/01 to 2021/01/01, and a test set spanning from 2021/01/01 to 2024/01/01. All price and volume data have been forward-dividend-adjusted respected to the adjustment factors on 2023/01/15.

To evaluate how well our framework performs against existing methods, tree models, heuristic algorithms, end-to-end deep model algorithms, and interpretable RL are adopted as baseline algorithms. We follow the open-source implementations of AlphaGen\cite{yu2023generating}, gplearn\cite{T_Stephens_2015} Stable Baseline 3\cite{raffin2021stable} and Qlib\cite{yang2020qlib} to produce the results.

\begin{itemize}
    \item Tree Model Algorithms: 
    \begin{itemize}
        \item XGBoost\cite{zhu2022application}: An efficient implementation of gradient boosting decision trees, which is known for its accuracy by combining multiple decision trees.
        \item LightGBM\cite{li2022research}: Another popular implementation of gradient boosting decision trees, which excels in speed and memory efficiency, making it ideal for quick analysis.
    \end{itemize}
    \item Heuristic Algorithms:
    \begin{itemize}
        \item GP\cite{zhang2020autoalpha}: A heuristic search algorithm for solving complex optimization problems, by generating and refining a population of candidate solutions.
    \end{itemize} 
    \item End-to-End Deep Model Algorithms: 
    \begin{itemize}
        \item MLP\cite{rumelhart1986learning}: A type of fully-connected feed-forward artificial neural network designed to learn complex patterns and relationships in the data
    \end{itemize}
    \item Interpretable Reinforcement Learning Algorithms: 
    \begin{itemize}
        \item AlphaGen\cite{yu2023generating}: A natural solution to the MDPs for mining formulaic alphas, utilizing RL for the first time in finding such interpretable alpha factors.
    \end{itemize}
\end{itemize}

In addition, both QFR and AlphaGen are used for solving the MDPs defined in Section \ref{Alpha Factor Mining Processes}. In order to demonstrate the performance of QFR in this RL task, some respected RL algorithms, including TRPO\cite{schulman2015trust}, PPO\cite{schulman2017proximal} and A3C\cite{ mnih2016asynchronous} are also utilized in the experiment.

\begin{figure*}[!ht]
\centering
\includegraphics[width=0.85\textwidth]{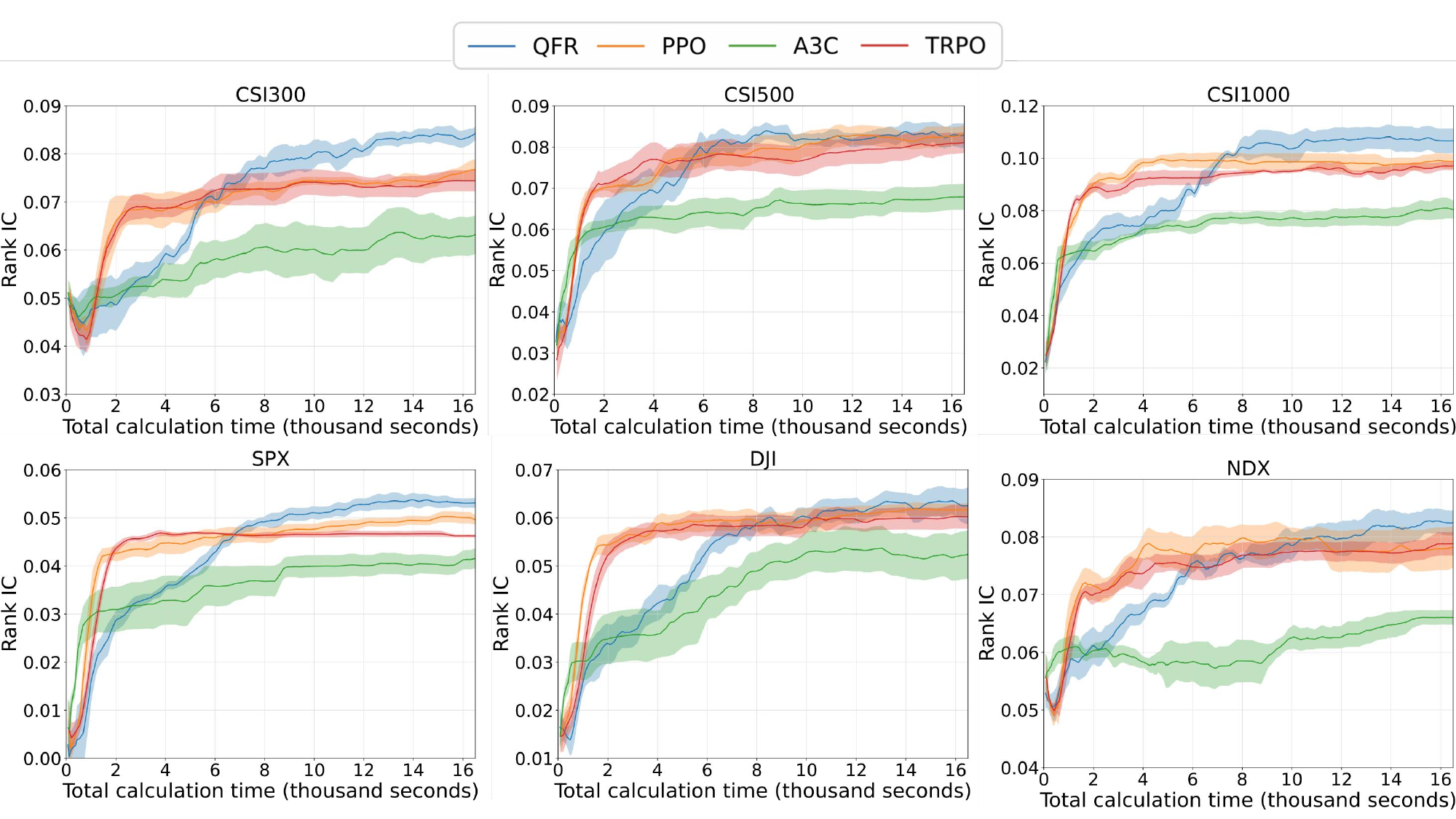}

\caption{Training-phase correlation between mined factor values and the prices of the six index constituents for the learning time of all investigated RL algorithms. All the curves are averaged over 5 different random seeds, and half of the standard deviation is shown as a shaded region.}
\label{Figure7}
\end{figure*}

To demonstrate the effect caused by stochasticity, each experiment is conducted with 5 different random seeds. The hyperparameters of MLP, XGBoost and LightGBM are set according to the benchmarks given by Qlib. The hyperparameters of GP are set according to the gplearn framework. The actor network and critical network of PPO, A3C and TRPO share a base LSTM feature extractor, which has a 2-layer structure with a hidden layer dimension of 128. The drop out rate is set to 0.1. The separate value and policy heads are MLPs with two hidden layers of 64 dimensions. PPO clipping range $\epsilon$ is set to 0.2. TRPO trust region constraint upper bounds $\delta$ is set to $1.1\cdot10^{-5}$. All the operator tokens used in our experiment are shown in Table \ref{All the Operator Tokens Used in the Experiment}. The experiments is performed by a single machine with an Intel Core i9-13900K CPU and two NVIDIA GeForce RTX 4090 GPUs.

\begin{table}[htbp]
\centering
\caption{All the Operator Tokens Used in the Experiment}
\begin{tabular*}{0.4\textwidth}{@{\extracolsep{\fill}}cc}
\toprule
Operator                              & Category      \\ \hline
Abs($x$)                              & Cross-Section \\
Log($x$)                              & Cross-Section \\ \hline
$x+y,x-y,x \times y, x/y$             & Cross-Section \\
Larger($x,y$), Smaller($x,y$)         & Cross-Section \\ \hline
Ref($x,l$)                            & Time-Series   \\
Mean($x,l$), Medium($x,l$) Sum($x,l$) & Time-Series   \\
Std($x,l$), Var($x,l$)                & Time-Series   \\
Max($x,l$), Min($x,l$)                & Time-Series   \\
Mad($x,l$)                            & Time-Series   \\
Delta($x,l$)                          & Time-Series   \\
WMA($x,l$), EMA($x,l$)                & Time-Series   \\ \hline
Cov($x,y,l$), Corr($x,y,l$)             & Time-Series   \\ \bottomrule
\end{tabular*}
\label{All the Operator Tokens Used in the Experiment}
\end{table}

\subsection{Comparisons with Other RLs}
\label{Comparisons with Other RLs}

We present the experimental results of state-of-the-art RL algorithms (including TRPO, PPO, and A3C) for mining formulaic factors on six indices, as shown in Fig. \ref{Figure7}. Due to the fact that QFR utilizes reward shaping, while the other algorithms do not, Rank Information Coefficient (Rank IC) rather than reward is used as a metric when comparing the learning curves. Rank IC is just the IC of ranked data, defined as $Rank IC\left(\mathbf{z}^{\prime}_t, \mathbf{y}_t\right) = IC\left(r(\mathbf{z}^{\prime}_t), r(\mathbf{y}_t\right))$, where $r(\cdot)$ is the ranking operator. This metric is the higher the better. The six indices are constructed from stocks traded on diverse exchanges, including the Shanghai Stock Exchange, the Shenzhen Stock Exchange, the Nasdaq Stock Market, and the New York Stock Exchange. In effect, these stocks are very different: the CSI300 constituents tend to be large, well-established companies; the CSI500 constituents mainly comprise mid-cap stocks; the CSI1000 constituents is dominated by small- or lower-cap firms; the NDX constituents feature technology-driven and innovative companies; the DJI constituents represent established industry leaders; and the SPX constituents cover a wide array of large-cap US corporations. These different constituent stocks may reflect varying levels of market stress in different periods. For instance, NDX constituents tend to exhibit higher market stress during tech downturns. According to Fig. \ref{Figure7}, apart from the experiment on CSI500 constituents where QFR and PPO exhibit similar performance, QFR outperforms all other algorithms in convergence speed and stability and improves by 3.83\% on the metric compared to the best-performing PPO algorithm adopted by AlphaGen on the remaining five index constituents. Specifically, QFR catches up to surpass initially superior of PPO/TRPO. The early advantage of PPO/TRPO is due to their controlled policy updates step size and therefore stability in update during cold-start phase. However, the restricted step sizes limit further improvements and make it harder to escape local optima. In contrast, without such restriction, QFR benefits from more accurate gradient estimations and ultimately surpasses PPO and TRPO after sufficient training. Considering that the training set from 2016/01/01 to 2020/01/01 spans four years of trading time and includes various stocks from six indices across four exchanges, encompassing various levels of market trading stress, this indicates that QFR also demonstrates superior performance under varying market trading stress. This enhanced performance can be attributed to QFR discarding elements associated with the critic network, utilizing a subtractive baseline, and implementing reasonable reward shaping.

\begin{figure*}[t]
\centering
\includegraphics[width=0.99\textwidth]{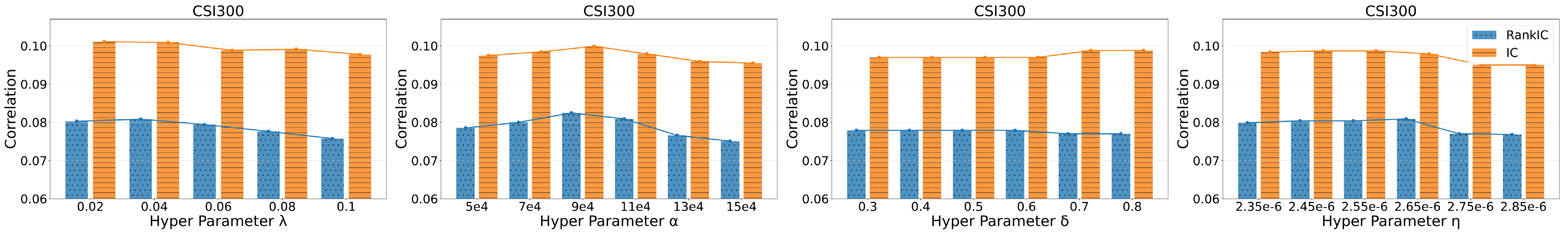}

\caption{
Sensitivity analysis of QFR to shaping coefficient ($\lambda$), time delay ($\alpha$), maximum IR test value ($\delta$), and slope ($\eta$).}

\label{Figure9}
\end{figure*}

\subsection{Sensitivity Analysis of the Time-varying Reward Shaping}
\label{The Studies of the Time-varying Reward Shaping}
We analyze the effects of reward shaping hyperparameters, including slope $\eta$, time delay $\alpha$, maximum IR test value $\delta$, and shaping coefficient $\lambda$. The dataset, by comprising the 300 largest and most liquid A-share stocks that collectively represent over 70\% of the market capitalization, provides a representative foundation for evaluations. As shown in Fig. \ref{Figure9}, increasing $\eta$ generally improves performance, but excessive values (e.g., $\eta>2.65 \times 10^{-6}$ ) intensify IR tests, forcing the policy network to receive prolonged discounted rewards after $\alpha$ steps. We select $\eta=2.65 \times 10^{-6}$ to balance IC and IR trade-offs. The time delay $\alpha$ decreases with enhanced factor quality, yet values below $7 \times 10^4$ trigger premature IR test and destabilize the process. Setting $\alpha=9 \times 10^4$ optimally activates IR test only after sufficient factor refinement. Notably, the maximum IR test value $\delta$ exhibits robustness, and we fix it at 0.3. For shaping coefficient $\lambda$, performance stays steady between $\lambda=0.02$ and $0.04$, but degrades significantly beyond $0.04$. For instance, $\lambda=0.06$ reduces IC by 2.27\% and Rank IC by 1.12\%, and $\lambda=0.1$ reduces IC by 3.36\% and Rank IC by 5.73\% compared to $\lambda=0.02$. While we explored adaptive $\lambda$ tuning, it ultimately underperformed the optimal static $\lambda=0.02$. We conservatively adopt $\lambda=0.02$ to avoid potential over‑amplification.

\subsection{Factors Evaluation}
\label{FacEva}
We applied the baseline algorithms and QFR to the CSI300 and CSI500 indices, and evaluated their performance based on two metrics: IC and Rank IC. The IC is delineated in (\ref{IC defination}). Both of the metrics are the higher the better. Note that all the algorithm is only optimized against  the IC metric. The results are shown in Table \ref{Performance of Mined Factors on CSI300 and CSI500}. The performance of MLP, XGBoost, and LightGBM is inferior due to the usage of the open-source set of alphas. On the other hand, AlphaGen tends to converge to local optima, leading to overfitting on the training data. Although GP mitigates this issue by preserving a diverse population, it still struggles to generate alphas that exhibit synergy in combination. QFR excels in solving MDPs for mining formulaic alpha factors compared to AlphaGen. We attribute the superior performance of QFR to the acceleration of the convergence process by discarding elements related to the critic network, as well as the reasonable reward shaping.
\begin{table}[tbh]
\centering
\caption{Testing-Phase Correlation Between the Prices of CSI300/CSI500 Constituents and Factor Values Mined by All Investigated Factor Mining Algorithms}
\label{Performance of Mined Factors on CSI300 and CSI500}
\begin{tabular*}{0.48\textwidth}{@{\extracolsep{\fill}}c|cc|cc}
\toprule
\multirow{2}{*}{}   & \multicolumn{2}{c|}{CSI300} & \multicolumn{2}{c}{CSI500}  \\ \cline{2-5} 
                          & IC           & RankIC        & IC  & RankIC      \\ \hline
\multirow{2}{*}{MLP}      & 0.0123                 & 0.0178                 & 0.0158       & 0.0211               \\
                          & (0.0006)               & (0.0017)      & (0.0014)     & (0.0007)             \\ \hdashline
\multirow{2}{*}{XGBoost}  & 0.0192                 & 0.0241                 & 0.0173       & 0.0217               \\
                          & (0.0021)               & (0.0027)               & (0.0017)     & (0.0022)             \\ \hdashline
\multirow{2}{*}{LightGBM} & 0.0158                 & 0.0235                 & 0.0112       & 0.0212               \\
                          & (0.0012)               & (0.0030)               & (0.0012)     & (0.0020)             \\ \hdashline
\multirow{2}{*}{GP}       & 0.0445                 & \textbf{0.0673 }       & \textbf{0.0557}  & 0.0665               \\
                          & (0.0044)               & \textbf{(0.0058)}      & \textbf{(0.0117)} & (0.0154)             \\ \hdashline
\multirow{2}{*}{AlphaGen} 
                          & \textbf{0.0500}        & 0.0540                & 0.0544       & \textbf{0.0722}               \\
                          & \textbf{(0.0021)}      & (0.0035)               & (0.0011)     & \textbf{(0.0017)}             \\ \hline
\multirow{2}{*}{QFR}    &\textbf{0.0588}  & \textbf{0.0602}        & \textbf{0.0708 }    &  \textbf{0.0674}     \\
                          &  \textbf{(0.0022)}      &  \textbf{(0.0014)}      &  \textbf{(0.0063) }   & \textbf{(0.0033)}  \\ \bottomrule
\end{tabular*}
\end{table}

\begin{table*}[!h]
\setlength{\tabcolsep}{5.8pt}
\centering
\caption{Testing-Phase Backtesting Market Conditions and Risk Metrics of the Investment Strategy Driven by QFR and AlphaGen for Each Quarter on CSI300 Constituents}
\begin{tabular}{
>{\columncolor[HTML]{FFFFFF}}c 
>{\columncolor[HTML]{FFFFFF}}c 
>{\columncolor[HTML]{FFFFFF}}c |
>{\columncolor[HTML]{FFFFFF}}c 
>{\columncolor[HTML]{FFFFFF}}c 
>{\columncolor[HTML]{FFFFFF}}c 
>{\columncolor[HTML]{FFFFFF}}c 
>{\columncolor[HTML]{FFFFFF}}c 
>{\columncolor[HTML]{FFFFFF}}c 
>{\columncolor[HTML]{FFFFFF}}c 
>{\columncolor[HTML]{FFFFFF}}c 
>{\columncolor[HTML]{FFFFFF}}c }
\toprule
\multicolumn{3}{l|}{\cellcolor[HTML]{FFFFFF}}                                                                                                                                                                                                    & 21Q1                                                                  & 21Q2                                                             & 21Q3                                                                            & 21Q4                                                    & 22Q1                                                         & 22Q2                                                             & 22Q3                                                            & 22Q4                                                               & 23Q1                                                                 \\ \hline
\multicolumn{3}{c|}{\cellcolor[HTML]{FFFFFF}Market Features}                                                                                                                                                                                       & \begin{tabular}[c]{@{}c@{}}Highest \\ Daily\\ Volatility\end{tabular} & \begin{tabular}[c]{@{}c@{}}Lowest \\ Daily\\ Volume\end{tabular} & \begin{tabular}[c]{@{}c@{}}Highest\\ Daily \\ Volume\\ \& Turnover\end{tabular} & \begin{tabular}[c]{@{}c@{}}Normal\\ Market\end{tabular} & \begin{tabular}[c]{@{}c@{}}Low\\ Bench\\ Return\end{tabular} & \begin{tabular}[c]{@{}c@{}}Highest\\ Bench\\ Return\end{tabular} & \begin{tabular}[c]{@{}c@{}}Lowest\\ Bench\\ Return\end{tabular} & \begin{tabular}[c]{@{}c@{}}Lowest \\ Daily\\ Turnover\end{tabular} & \begin{tabular}[c]{@{}c@{}}Lowest \\ Daily\\ Volatility\end{tabular} \\ \hline
\multicolumn{1}{c|}{\cellcolor[HTML]{FFFFFF}}                                                                                 & \multicolumn{2}{c|}{\cellcolor[HTML]{FFFFFF}Daily Volatility}                                                    & \textbf{22.79\%}                                                      & 21.06\%                                                          & 19.68\%                                                                         & 17.94\%                                                 & 19.19\%                                                      & 21.71\%                                                          & 19.08\%                                                         & 20.68\%                                                            & \textbf{17.90\%}                                                     \\
\multicolumn{1}{c|}{\cellcolor[HTML]{FFFFFF}}                                                                                 & \multicolumn{2}{c|}{\cellcolor[HTML]{FFFFFF}(CIMV)}                                                              & \textbf{(1.58\%)}                                                     & \multicolumn{1}{l}{\cellcolor[HTML]{FFFFFF}(1.08\%)}             & (1.25\%)                                                                        & \multicolumn{1}{l}{\cellcolor[HTML]{FFFFFF}(1.27\%)}    & \multicolumn{1}{l}{\cellcolor[HTML]{FFFFFF}(2.63\%)}         & \multicolumn{1}{l}{\cellcolor[HTML]{FFFFFF}(1.71\%)}             & \multicolumn{1}{l}{\cellcolor[HTML]{FFFFFF}(1.69\%)}            & \multicolumn{1}{l}{\cellcolor[HTML]{FFFFFF}(1.93\%)}               & \textbf{(1.06\%)}                                                             \\ \cline{2-12} 
\multicolumn{1}{c|}{\cellcolor[HTML]{FFFFFF}}                                                                                 & \multicolumn{2}{c|}{\cellcolor[HTML]{FFFFFF}Daily Volume}                                                        & 6.96                                                                  & \textbf{6.70}                                                    & \textbf{9.13}                                                                   & 7.85                                                    & 8.11                                                         & 8.31                                                             & 7.31                                                            & 6.93                                                               & 6.94                                                                 \\
\multicolumn{1}{c|}{\cellcolor[HTML]{FFFFFF}}                                                                                 & \multicolumn{2}{c|}{\cellcolor[HTML]{FFFFFF}(100 Million Lots)}                                         & (0.89)                                                      & \textbf{(0.64)}                                                  & \textbf{(1.48)}                                                                 & (0.72)                                         & (1.03)                                              & (0.88)                                                  & (1.04)                                                 & (1.14)                                                & (0.90)                                                      \\ \cline{2-12} 
\multicolumn{1}{c|}{\cellcolor[HTML]{FFFFFF}}                                                                                 & \multicolumn{2}{c|}{\cellcolor[HTML]{FFFFFF}Daily Turnover}                                                      & 0.94                                                                  & 0.88                                                             & \textbf{1.30}                                                                   & 1.10                                                    & 1.01                                                         & 0.95                                                             & 0.91                                                            & \textbf{0.84}                                                      & 0.88                                                                 \\
\multicolumn{1}{c|}{\cellcolor[HTML]{FFFFFF}}                                                                                 & \multicolumn{2}{c|}{\cellcolor[HTML]{FFFFFF}(Trillion Yuan)}                                                     & (0.16)                                                                & (0.12)                                                           & \textbf{(0.16)}                                                                          & (0.09)                                                  & (0.13)                                                       & (0.15)                                                           & (0.16)                                                          & \textbf{(0.15)}                                                             & (0.12)                                                               \\ \cline{2-12} 
\multicolumn{1}{c|}{\multirow{-7}{*}{\cellcolor[HTML]{FFFFFF}\begin{tabular}[c]{@{}c@{}}Market\\ Conditions\end{tabular}}}    & \multicolumn{2}{c|}{\cellcolor[HTML]{FFFFFF}Bench Return}                                                        & -3.13\%                                                               & 1.21\%                                                           & -6.97\%                                                                         & 1.52\%                                                  & -14.53\%                                                     & \textbf{6.21\%}                                                  & \textbf{-15.16\%}                                               & 1.75\%                                                             & 4.26\%                                                               \\ \hline
\multicolumn{1}{c|}{\cellcolor[HTML]{FFFFFF}}                                                                                 & \cellcolor[HTML]{FFFFFF}                                                                              & QFR      & 7.55\%                                                                & -1.80\%                                                          &  21.06\%                                                                         & 1.59\%                                                  & 0.48\%                                                       & 2.65\%                                                           & -6.08\%                                                         & 9.10\%                                                             & 9.74\%                                                               \\
\multicolumn{1}{c|}{\cellcolor[HTML]{FFFFFF}}                                                                                 & \multirow{-2}{*}{\cellcolor[HTML]{FFFFFF}\begin{tabular}[c]{@{}c@{}}Cumulative\\ Return\end{tabular}} & AlphaGen & 4.55\%                                                                & 9.48\%                                                           & 20.97\%                                                                         & 0.21\%                                                  & -6.47\%                                                      & 8.30\%                                                           & -6.78\%                                                         & 2.09\%                                                             & 7.46\%                                                               \\ \cline{2-12} 
\multicolumn{1}{c|}{\cellcolor[HTML]{FFFFFF}}                                                                                 & \cellcolor[HTML]{FFFFFF}                                                                              & QFR      & 1.912                                                                 & -0.170                                                           & 2.124                                                                           & 0.193                                                   & 0.227                                                        & 0.445                                                            & -0.689                                                          & 1.201                                                              & 1.405                                                                \\
\multicolumn{1}{c|}{\cellcolor[HTML]{FFFFFF}}                                                                                 & \multirow{-2}{*}{\cellcolor[HTML]{FFFFFF}\begin{tabular}[c]{@{}c@{}}Sharpe\\ Ratio\end{tabular}}      & AlphaGen & 1.116                                                                 & 2.700                                                            & 3.432                                                                           & 0.106                                                   & -1.145                                                       & 1.405                                                            & -1.560                                                          & 0.553                                                              & 2.418                                                                \\ \cline{2-12} 
\multicolumn{1}{c|}{\cellcolor[HTML]{FFFFFF}}                                                                                 & \cellcolor[HTML]{FFFFFF}                                                                              & QFR      & 5.88\%                                                                & 3.99\%                                                           & 4.60\%                                                                          & 7.05\%                                                  & 10.33\%                                                      & 8.55\%                                                           & 7.80\%                                                          & 7.27\%                                                             & 4.38\%                                                               \\
\multicolumn{1}{c|}{\cellcolor[HTML]{FFFFFF}}                                                                                 & \multirow{-2}{*}{\cellcolor[HTML]{FFFFFF}\begin{tabular}[c]{@{}c@{}}Maximum\\ Drawdown\end{tabular}}  & AlphaGen & 5.10\%                                                                & 3.72\%                                                           & 5.08\%                                                                          & 7.14\%                                                  & 8.73\%                                                       & 8.16\%                                                           & 7.88\%                                                          & 6.56\%                                                             & 3.41\%                                                               \\ \cline{2-12} 
\multicolumn{1}{c|}{\cellcolor[HTML]{FFFFFF}}                                                                                 & \cellcolor[HTML]{FFFFFF}                                                                              & QFR      & 183.71\%                                                              & 129.66\%                                                         & 605.62\%                                                                        & 327.65\%                                                & 439.15\%                                                     & 531.74\%                                                         & 114.90\%                                                        & 148.96\%                                                           & 79.94\%                                                              \\
\multicolumn{1}{c|}{\multirow{-8}{*}{\cellcolor[HTML]{FFFFFF}\begin{tabular}[c]{@{}c@{}}Strategy\\ Risk\\Metrics\end{tabular}}} & \multirow{-2}{*}{\cellcolor[HTML]{FFFFFF}Turnover}                                                    & AlphaGen & \multicolumn{1}{l}{\cellcolor[HTML]{FFFFFF}446.55\%}                  & \multicolumn{1}{l}{\cellcolor[HTML]{FFFFFF}508.80\%}             & \multicolumn{1}{c}{\cellcolor[HTML]{FFFFFF}452.97\%}                            & \multicolumn{1}{l}{\cellcolor[HTML]{FFFFFF}321.84\%}    & \multicolumn{1}{l}{\cellcolor[HTML]{FFFFFF}347.68\%}         & \multicolumn{1}{l}{\cellcolor[HTML]{FFFFFF}385.03\%}             & \multicolumn{1}{l}{\cellcolor[HTML]{FFFFFF}396.92\%}            & \multicolumn{1}{l}{\cellcolor[HTML]{FFFFFF}445.46\%}               & \multicolumn{1}{l}{\cellcolor[HTML]{FFFFFF}482.29\%}                 \\ \bottomrule
\end{tabular}
\label{Risk Metrics}
\end{table*}

\subsection{Investment Simulation}
\label{Investment Simulation}
\begin{figure}[!h]
\centering
\includegraphics[width=0.45\textwidth]{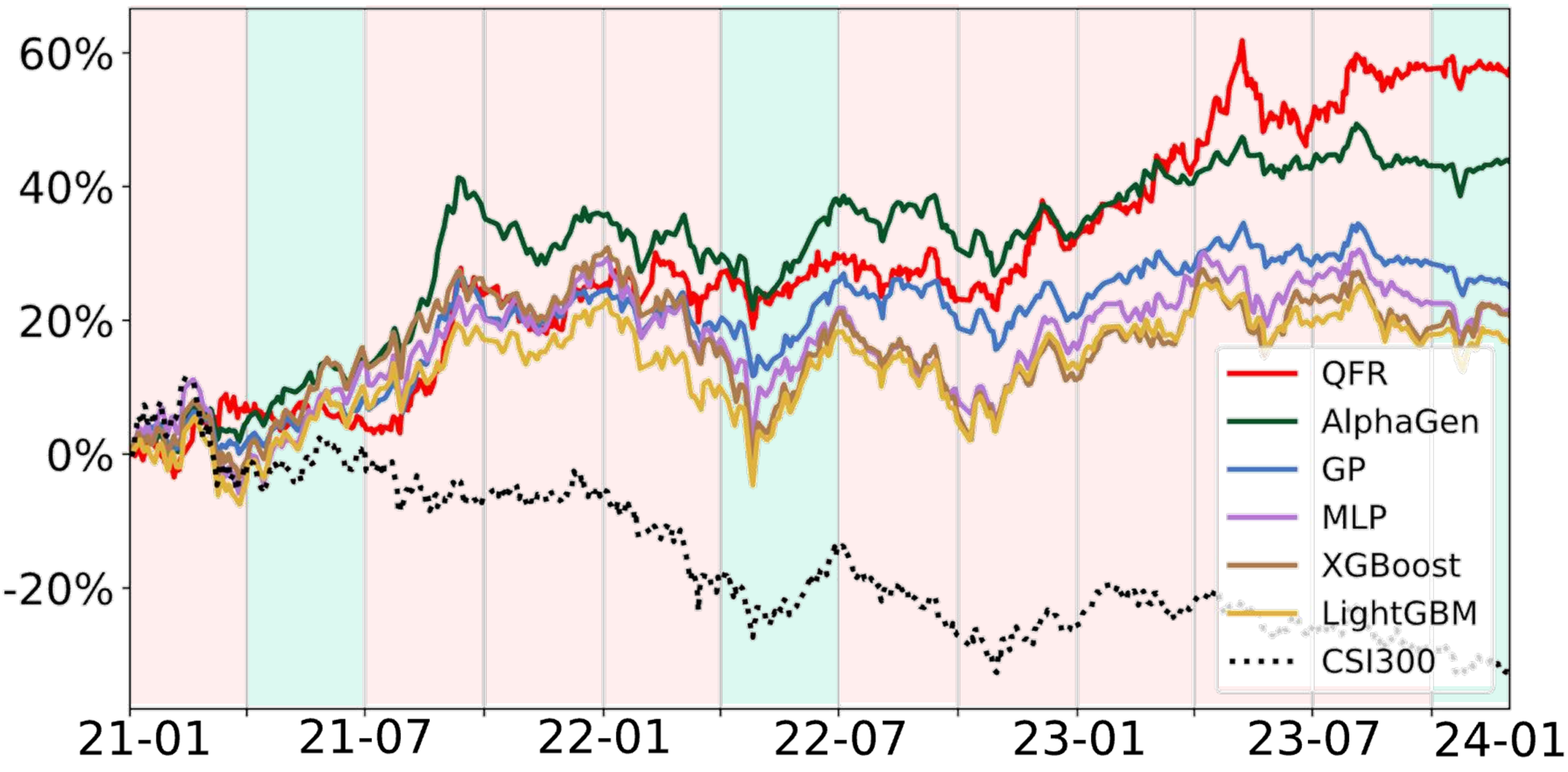}
\caption{Testing-phase backtesting results of the investment strategies driven by all investigated factor mining algorithms on CSI300 constituents. The lines track the cumulative returns of simulated investment strategies using the mined factors. The red background indicates that QFR dominates the compared methods with quarterly cumulative returns, while a green background indicates the contrary.}
\label{Figure5}
\end{figure}

To further demonstrate QFR's effectiveness in realistic investments, we conduct simulations using strategies driven by all investigated algorithm. The testing phase(2021/01/01–2024/01/01) employs CSI300 index, which tracks the 300 largest and most liquid A-share stocks. As China’s premier equity benchmark, it captures over 70\% of the market’s total value, ensuring representative evaluations. The trading strategy is a top-50 strategy. On each trading day, the CSI300 constituents are first sorted according to the mined alpha factor values, then the top-50 stocks are selected for holding positions. If the top-50 stock list changes compared to the previous trading day, rebalancing is required (buying new entries and selling those removed), and we assume the difference between the expected and actual execution price is zero because daily trading using TWAP/VWAP strategies effectively neutralizes this kind of price slippage\cite{slippage,HFT,vwap,wang2024alleviating,li2024simlob}. The performance of the investment strategy was evaluated using cumulative returns, with the final value of this metric being the higher, the better. The result of the backtest is shown in Fig. \ref{Figure5}, Table \ref{Risk Metrics} and Table \ref{Time Horizons}. 

\begin{figure*}[!h]
\centering
\includegraphics[width=0.85\textwidth]{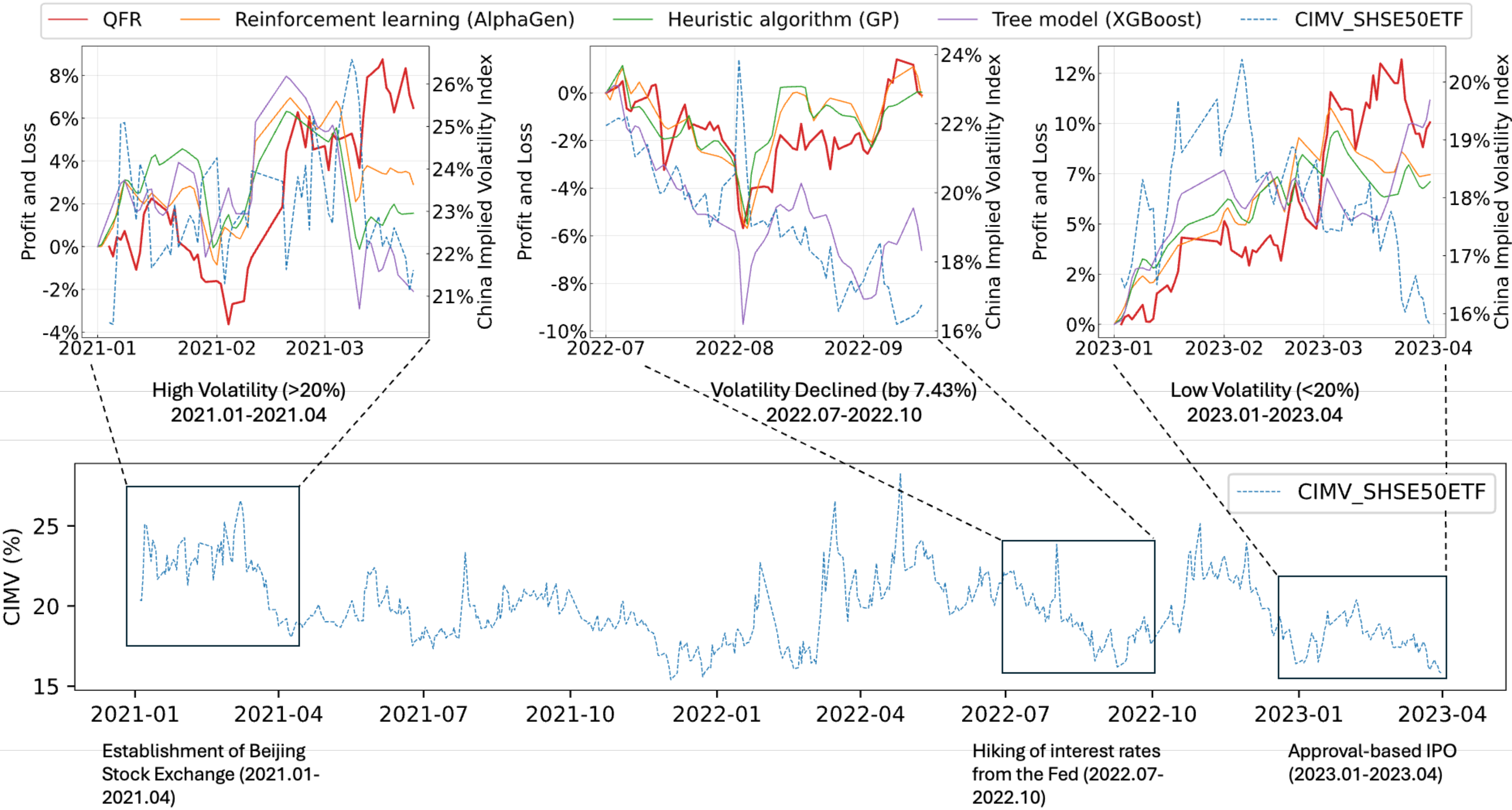}

\caption{Testing-phase backtesting performance of the investment strategies driven by all investigated factor mining algorithms on CSI300 constituents under different market volatility conditions. Three significant event-driven volatility anomaly periods were selected to demonstrate the ability of each algorithm to withstand volatility.  
}

\label{Figure6}
\end{figure*}

\begin{figure*}[!h]
\centering
\includegraphics[width=0.85\textwidth]{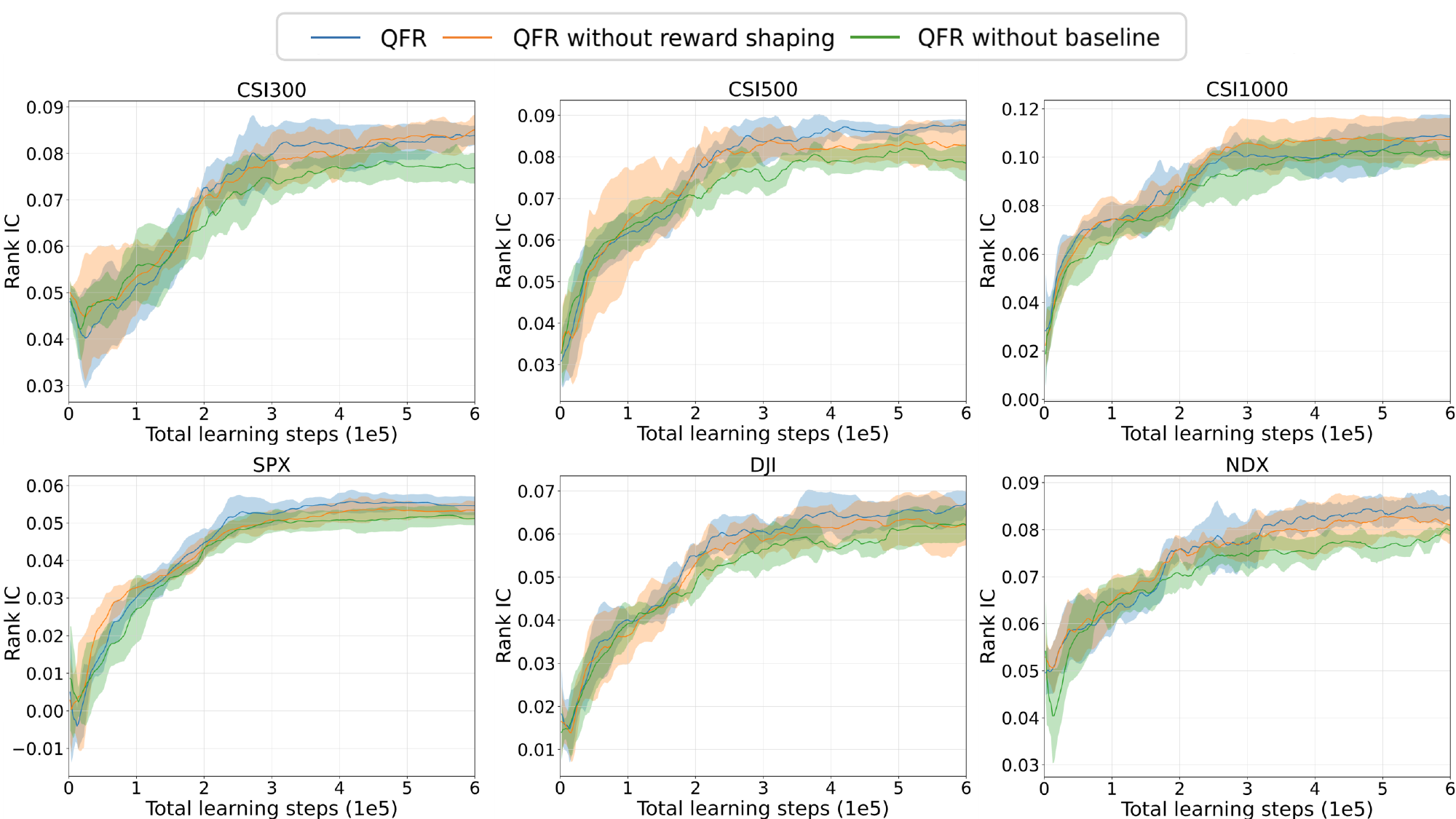}

\caption{Training-phase correlation between mined factor values and the prices of the six index constituents of  for each learning episode of the two variants, together with QFR. All the curves in this ablation experiments are averaged over 5 different random seeds, and half of the standard deviation is shown as a shaded region.}
\label{Figure10}
\end{figure*}

It can be observed that QFR outperforms the baseline methods under most market conditions. Among 12 backtested quarters, the QFR-driven investment strategy failed to dominate in only 3 quarters. Even in periods of highest market daily volatility (21Q1), highest market daily volume and turnover (21Q3), lowest benchmark return (22Q3), lowest market daily turnover (22Q4), and lowest market daily volatility (23Q1), the QFR-driven investment strategy still outperforms the best baseline algorithm, AlphaGen, demonstrating its robust effectiveness under various market conditions and market stress. In the cumulative return comparisons across weekly, quarterly, and yearly time horizons, the QFR-driven investment strategy still outperforms all baseline algorithms, only slightly underperforming relative to AlphaGen in the monthly cumulative returns comparison, demonstrating its robust performance and consistent superiority over multiple time horizons.

\begin{table}[!t]
\centering
\caption{Testing-Phase Cumulative Returns of the Investment Strategies Driven by All Investigated Factor Mining Algorithms on CSI300 Constituents under Different Time Horizons }
\begin{tabular*}{0.48\textwidth}{@{\extracolsep{\fill}}c|cccc}
\toprule
\multirow{2}{*}{} & \multicolumn{4}{c}{Cumulative Returns by Time Horizons (\%)}                                                                                                                                                                                              \\ \cline{2-5} 
                  & Weekly                                                  & Monthly                                                 & Quarterly                                               & Yearly                                                    \\ \hline
MLP               & \begin{tabular}[c]{@{}c@{}}0.124\\ (2.374)\end{tabular} & \begin{tabular}[c]{@{}c@{}}0.387\\ (5.478)\end{tabular} & \begin{tabular}[c]{@{}c@{}}1.733\\ (8.770)\end{tabular} & \begin{tabular}[c]{@{}c@{}}7.158\\ (20.603)\end{tabular}  \\ \hdashline
XGBoost           & \begin{tabular}[c]{@{}c@{}}0.128\\ (2.226)\end{tabular} & \begin{tabular}[c]{@{}c@{}}0.423\\ (5.459)\end{tabular} & \begin{tabular}[c]{@{}c@{}}1.702\\ (9.514)\end{tabular} & \begin{tabular}[c]{@{}c@{}}6.790\\ (24.623)\end{tabular}  \\ \hdashline
LightGBM          & \begin{tabular}[c]{@{}c@{}}0.104\\ (2.243)\end{tabular} & \begin{tabular}[c]{@{}c@{}}0.333\\ (4.658)\end{tabular} & \begin{tabular}[c]{@{}c@{}}1.402\\ (8.629)\end{tabular} & \begin{tabular}[c]{@{}c@{}}5.538\\ (15.901)\end{tabular}  \\ \hdashline
GP                & \begin{tabular}[c]{@{}c@{}}0.127\\ (1.679)\end{tabular} & \begin{tabular}[c]{@{}c@{}}0.636\\ (4.269)\end{tabular} & \begin{tabular}[c]{@{}c@{}}2.115\\ (5.584)\end{tabular} & \begin{tabular}[c]{@{}c@{}}8.292\\ (14.562)\end{tabular}  \\ \hdashline
AlphaGen          & \begin{tabular}[c]{@{}c@{}}0.238\\ (1.875)\end{tabular} & \begin{tabular}[c]{@{}c@{}}\textbf{1.114}\\ \textbf{(5.002)}\end{tabular} & \begin{tabular}[c]{@{}c@{}}3.620\\ (7.814)\end{tabular} & \begin{tabular}[c]{@{}c@{}}14.515\\ (19.579)\end{tabular} \\ \hline
QFR               & \begin{tabular}[c]{@{}c@{}}\textbf{0.247}\\ \textbf{(2.057)}\end{tabular}   & \begin{tabular}[c]{@{}c@{}}0.999\\ (4.341)\end{tabular} & \begin{tabular}[c]{@{}c@{}}\textbf{4.277}\\ \textbf{(6.806)}\end{tabular} & \begin{tabular}[c]{@{}c@{}}\textbf{18.673}\\ \textbf{(10.541)}\end{tabular} \\ \bottomrule
\end{tabular*}
\label{Time Horizons}
\end{table}

Additionally, based on the China Implied Volatility Index (CIMV) \cite{CIMV}, periods of high volatility (2021/01/01-2021/03/24, with CIMV $> 20\%$), periods of rapid volatility changes (2022/07/01-2022/09/15, with a decrease in CIMV of $7.43\%$), and periods of low volatility (2023/01/01-2023/03/31, with CIMV $< 20\%$) are selected to further demonstrate the factor's ability to withstand volatility. We then analyzed the profit and loss for each period, and the results are shown in Fig. \ref{Figure6}. Our algorithm was able to achieve the most profit under various volatility conditions. In particular, during periods of high volatility, QFR had a significant advantage over the baseline algorithms.

\subsection{Ablation Study}
\label{Ablation Study}




To investigate the role of the two improvements of QFR, we designed two variants including: QFR without baseline represents that the policy network is updated directly using REINFORCE; QFR without reward shaping represents that IC is used directly as reward without any IR test. Fig. \ref{Figure10} presents the ablation results. QFR can provide prominent performance among all six index constituent stocks, indicating that all improvements play important roles. When removing the baseline leads to a degradation in the quality of the factors, it implies that the baseline can help reduce the variance. While QFR without reward shaping learns faster at the beginning, it finally falls to local optima, which may be due to an overemphasis on the absolute returns. In summary, QFR conditioned on all improvements gives the best performance, indicating that baseline and reward shaping are complementary parts of QFR.

\section{Conclusion}
In this paper, we have proposed a novel RL algorithm, QuantFactor REINFORCE (QFR), for mining formulaic alpha factors. QFR leverages the advantages of discarding the critic network while theoretically addressing its limitations of high variance by introducing a greedy baseline. Additionally, the incorporation of IR as a reward shaping mechanism encourages the generation of stable alpha factors that can better adapt to changing market conditions. Our extensive experiments on real-world asset datasets demonstrate QFR achieves
better performance compared to other state-of-the-art RL algorithm when mining formulaic alpha factors. It also generates superior alpha factors over existing factor mining methods. We conclude that QFR is a promising approach for mining formulaic alpha factors. Beyond financial markets, the symbolic regression capabilities of QFR could be extended to diverse signal processing tasks that require interpretable, formula-based models, such as symbolic regression for time-frequency analysis\cite{2021Serial}, feature extraction in image denoising\cite{2023Deep}, and adaptive filtering in speech processing\cite{2018A}. Future research directions include integrating matrix tri-factorization techniques to enhance QFR's capability in capturing cross-stock correlations and leveraging community detection techniques to uncover latent asset clusters and investor behavior patterns.


\bibliographystyle{IEEEtran}
\bibliography{QFR.bib}

\end{document}